\documentclass[reprint, aps, prx, amsmath, amssymb, superscriptaddress, longbibliography]{revtex4-2}
\usepackage{bm}
\usepackage{graphicx}
\usepackage{comment}
\usepackage{float}
\usepackage{multirow}
\usepackage{bbding}
\usepackage[colorlinks, linkcolor= blue, citecolor = blue, urlcolor=blue]{hyperref}

\usepackage{xr}
\newcommand*{\addFileDependency}[1]{\typeout{(#1)}

\IfFileExists{#1}{}{\typeout{No file #1.}}
}\makeatother

\newcommand*{\myexternaldocument}[1]{%
\externaldocument{#1}%
}
\def\be{\begin{equation}}
\def\ee{\end{equation}}
\def \bea{\begin{eqnarray}}
\def \eea{\end{eqnarray}}
\def \nn{\nonumber}
\def\c{\CheckmarkBold}
\def\x{\XSolidBrush}
\myexternaldocument{SM_int_third_order}

\begin{document}
\title{Quantum Geometry Induced Third Order Nonlinear Transport Responses} 
\author{Debottam Mandal}
\thanks{D.M. and S.S. contributed equally and are the joint first authors.}
\affiliation{Department of Physics, Indian Institute of Technology Kanpur, Kanpur 208016, India}
\author{Sanjay Sarkar}
\thanks{D.M. and S.S. contributed equally and are the joint first authors.}
\affiliation{Department of Physics, Indian Institute of Technology Kanpur, Kanpur 208016, India}
\author{Kamal Das}
\email{daskamal457@gmail.com}
\affiliation{Department of Condensed Matter Physics, Weizmann Institute of Science, Rehovot 7610001, Israel}
\author{Amit Agarwal}
\email{amitag@iitk.ac.in}
\affiliation{Department of Physics, Indian Institute of Technology Kanpur, Kanpur 208016, India}

\begin{abstract}

Nonlinear transport phenomena offer an exciting probe into the band geometry and symmetry properties of a system. While most studies on nonlinear transport have looked at second-order nonreciprocal responses in noncentrosymmetric systems, the reciprocal third-order effects dominant in centrosymmetric systems remain largely uncharted. Here, we uncover two significant contributions to third-order charge conductivity: one affecting longitudinal resistance and another impacting the Hall effect. We demonstrate that these previously unexplored contributions arise in time-reversal symmetry-broken systems from band geometric quantities such as the Berry curvature and the symplectic connection. We prescribe a detailed symmetry dictionary to facilitate the discovery of these fundamental transport coefficients.  Additionally, we unify our quantum kinetic results with the semiclassical wave-packet formalism to unveil all contributions to third-order charge transport. We illustrate our results in antiferromagnetic monolayer SrMnBi$_2$. Our comprehensive study significantly advances the fundamental understanding of reciprocal nonlinear responses. 
 
\end{abstract}

\maketitle

\section{Introduction}

In recent years, the study of nonlinear transport phenomena in quantum materials has attaracted significant interest~\cite{Du_nphyreview2021_nonlinear,Ortix2021,rikken_PRL2001_electric,Deyo2009,Moore2010,YGao2014,Sodemann2015,Facio2018,QMa2019,Kang2019,Du2019,Wang_SciAdv2019_ferro_photocurrent,He_PRL2019_nphe,shao_PRL2020_nonlin,Xiao_Natcom2020_memory,Isobe_SciAdv2020_highfrequency,Kumar_Natnano2021_rf_rectification,HLiu2021_intrinsic,Tiwari_Natcom2021_giantcaxis,Zhang_PNAS2021_terahertz,Wang2021_intrinsic,Huang_PRL2022_intrinsic_nphe,Chakraborty_2022,chen_PRL2022_role,oiwa_JPSJ2022_sys,Lahiri2022_resistivity,Das2023_intrinsic,Wei_PRL2023_quanfluctuation,Zhang2023_symmetry,Gao2023_science,wang_Nat2023_qun,Kaplan2023_unification,Kaplan_Natcom2023_general_nh,Wang_PRL2023_neelafm,Bandyopadhyay_2024arxiv_nonlinear}. This surge is driven by both potential  applications~\cite{Xiao_Natcom2020_memory,Isobe_SciAdv2020_highfrequency,Kumar_Natnano2021_rf_rectification, Zhang_PNAS2021_terahertz} and the opportunity to discover new  physical and material properties beyond the linear response theories. These studies have unveiled significant insights on the quantum geometry of the Bloch bands~\cite{Provost1980,Morimoto2016a,Nagaosa2017,Ahn2021,Ma2021_spotlight,Torma2023_essay_quant_geo,Deyo2009,Moore2010,Sodemann2015,Du2019,Ortix2021,QMa2019,Kang2019}, crystalline symmetries~\cite{Sodemann2015,Zhang2023_symmetry}, topological phase transitions~\cite{Sinha2022,Chakraborty_2022}, and magnetic states~\cite{YGao2014,HLiu2021_intrinsic,Wang2021_intrinsic, Das2023_intrinsic,shao_PRL2020_nonlin,chen_PRL2022_role}. However, most of these works have focused on non-reciprocal second order transport in inversion broken systems~\cite{tokura_NC2018_non,Sipe2000,Young2012,Král2000,Tan2016,deJuan2017,Isobe2020,Le2021,Singh_PRB_2018, Orenstein2021,nagaosa_AP2022_nonlinear, YGao2021_highharmonic,Bhalla2022,Bhalla2023}, 
and our understanding of nonlinear responses in centrosymmetric systems remains limited. 
To address this gap, there have been some recent efforts towards investigating the third-order reciprocal transport~\cite{Lai2021,Wang2022,Ye2022_orbital,li_arxiv2023_quantum,Zhao_PRL2023_bcd_polarizability}. 
Although initial studies have begun to chart this territory by 
examining specific contributions~\cite{Lai2021,HLiu2022_BCP_prb, Roy2022,Nag2022,Wang2022,Ye2022_orbital,Pal2023_polarization,zhu_PRB2023_third,Zhang2023_multipole,sankar_arxiv2023_observation, Sorn2023_quadrupole,Xiang2023,Fang2023_altermagnet}, the complete landscape for third-order reciprocal transport responses remains elusive. 

\begin{figure}[t!]
    \includegraphics[width=\linewidth]{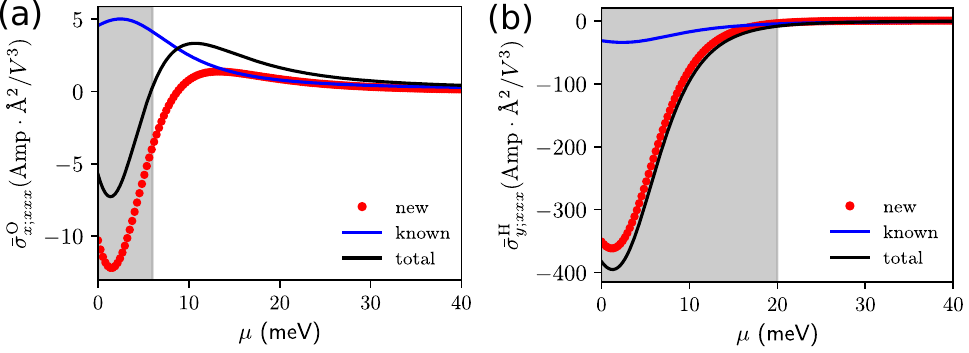}
    \caption{The variation of the third order nonlinear (a) longitudinal and (b) Hall conductivity of the antiferromagnetic monolayer of SrMnBi$_2$ with chemical potential ($\mu$). 
    The previously unexplored contributions are marked as `new'. 
    The shaded region in both plots highlights the regime where the new contributions dominate the known response. See Sec.~\ref{model_calc_tilted_warped} for details of the model calculations and parameters.}
\label{fig_00}
\end{figure}

Motivated by this, we predict two new significant contributions (see Fig.~\ref{fig_00}), and provide the complete picture of third-order nonlinear transport. 
%
%
%
Specifically, we predict a new and unique contribution to the third-order dissipative conductivity, which enables nonlinear, reciprocal longitudinal resistivity in inversion-symmetric systems. Additionally, we predict a novel and significant contribution to the third-order dissipationless Hall effect. We show that these previously unexplored contributions  dominate the third order nonlinear transport near the band edges, see  Fig.~\ref{fig_00} (shaded region). Consequently, these contributions are crucial for understanding and correctly interpreting recent third-order transport experiments~\cite{Lai2021, Wang2022,Ye2022_orbital,sankar_arxiv2023_observation}. Interestingly, both these contributions are independent of scattering time (intrinsic) and driven by the quantum geometry of Bloch electrons. 

\begin{figure}[t!]
    \includegraphics[width=\linewidth]{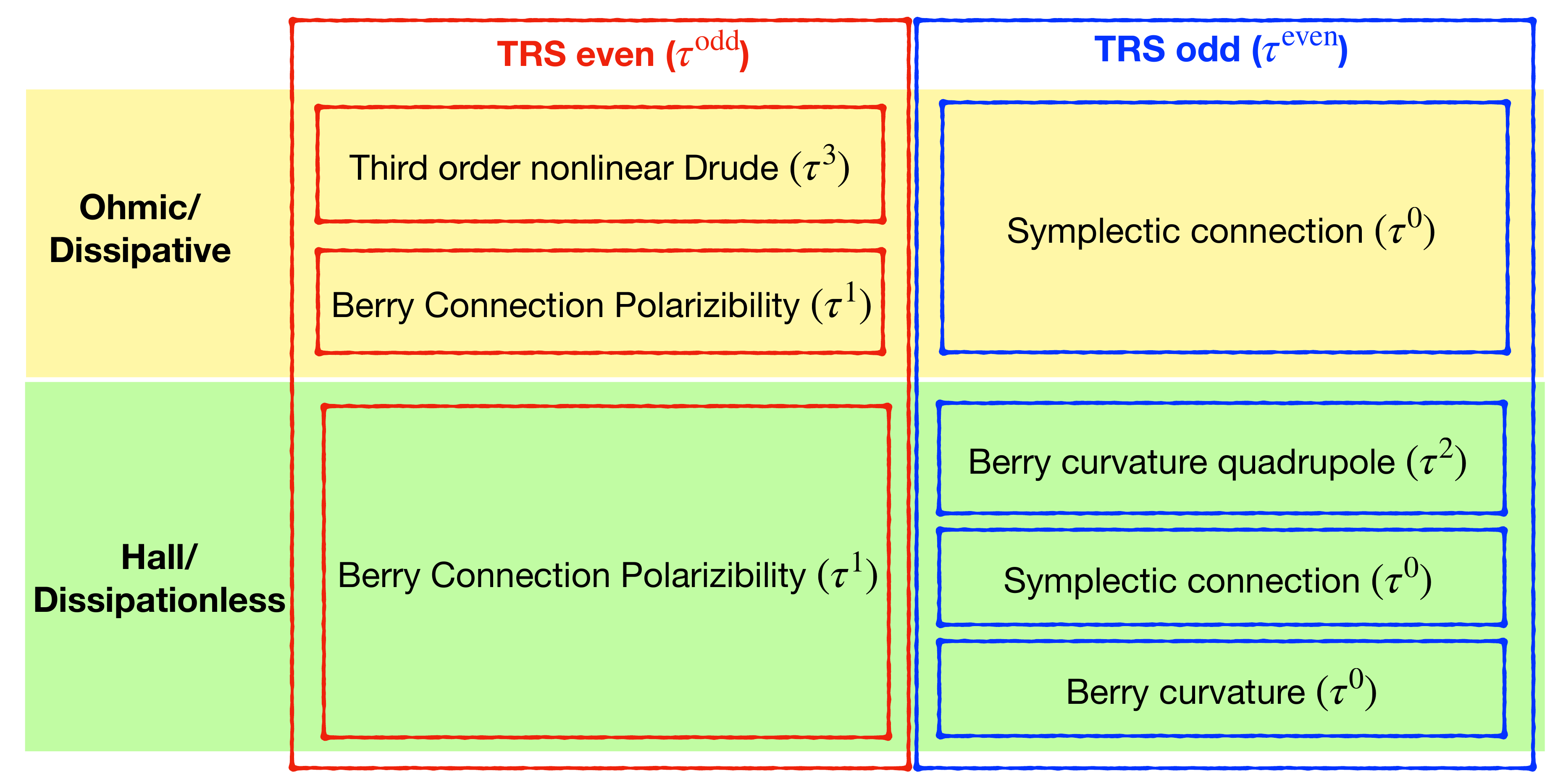}
    \caption{Schematic tabulating all the third-order transport coefficients.  We divide them into odd and even powers of scattering time. The even (odd) power contributions do (do not) require time-reversal breaking. Among the odd scattering time-dependent contributions ($\propto \tau^{\rm odd}$), the cubic scattering time-dependent contribution is dissipative. The linear scattering time-dependent contribution has both dissipative and Hall parts. Both are related to the Berry connection polarizability~\cite{Lai2021,HLiu2022_BCP_prb}.  Among the even power dependent contributions ($\propto \tau^{\rm even}$), the quadratic power contribution is completely Hall-like ~\cite{Parker2019,Zhang2023_multipole}. 
The symplectic connection-induced intrinsic Hall contribution was predicted in Ref.~\cite{Xiang2023}. We complete this table and predict two new contributions: the symplectic connection-induced dissipative response and the Berry curvature-induced intrinsic Hall response.
\label{fig_0}}
\end{figure}

To facilitate the experimental discovery of these contributions, we elucidate the crystalline and magnetic point group of materials that can host these contributions. Additionally, we reproduce all the previously explored contributions based on the wave-packet approach, from our quantum kinetic theory calculations. Together with the new contributions, this completes the theory of third-order response as highlighted in Fig.~\ref{fig_0}. Our work unifies the quantum kinetic theory framework with the semiclassical wave-packet theory for third-order responses and highlights several subtleties of calculations and interpretation. 
We demonstrate the novel intrinsic responses in an antiferromagnetic monolayer of SrMnBi$_{2}$ and discuss the experimental feasibility of observing the predicted third-order responses. Our comprehensive understanding of third-order responses will enable the exploration of the large class of centrosymmetric systems for potential applications based on nonlinear transport.

The rest of our paper is organized as follows. In Sec.~\ref{formalism}, we introduce the density matrix-based quantum kinetic theory---perturbative calculation of the density matrix up to third order in the electric field. In Sec.~\ref{third_order_cond}, we present the general expression of third-order currents and extract the conductivities. In Sec.~\ref{unifying_semiclasscs}, we connect and compare our results from the quantum kinetic theory with the previously known results of semiclassical wave-packet formalism. We present a detailed crystalline symmetry analysis of the third-order conductivities in Sec.~\ref{symmetry} to unveil the magnetic point groups that can host the third-order responses. In Sec.~\ref{model_calc_tilted_warped}, we illustrate the existence of the intrinsic conductivities predicted in this paper in an antiferromagnetic monolayer. This is followed by a scaling law analysis of the third-order conductivities in Sec.~\ref{scaling_law}. Finally, we conclude with a summary and outlook of our work in Sec.~\ref{conclude}, following a discussion in Sec.~\ref{discuss}.

\section{Density matrix formalism for nonlinear conductivities} \label{formalism}

We begin by introducing the key ingredients of the density matrix-based quantum kinetic framework that we will use to calculate the third-order conductivities. Specifically, we summarize the gauge convention of the electric field coupling to the charge carriers and perturbative solutions of the density matrix.

\subsection{Length gauge}
A material subjected to an electromagnetic field can be described by the Hamiltonian, ${\cal H} = \hat{H}_0 + \hat{H}_{\rm E}$, with $\hat{H}_0$ describing the unperturbed Bloch Hamiltonian and 
$\hat{H}_{\rm E}$ is the perturbation Hamiltonian capturing the effect of the electric field. There are two ways to incorporate the effect of the electric field: the velocity gauge~[\onlinecite{Moss1990,Sipe1993}] and the length gauge~[\onlinecite{Aversa1995}]. The Hamiltonians in these two approaches are related through a time-dependent unitary transformation and yield the same results in the response functions~[\onlinecite{Ventura2017,Passos2018}]. In our paper, we adopt the length gauge. In the dipole approximation, the interaction Hamiltonian is expressed as 
\be
\hat{H}_{\rm E} = e{\bm E}(t) \cdot \hat{\bm r} ~.
\ee
Here, $-e$ is the electronic charge, $\hat{\bm r}$ is the position operator, and ${\bm E}(t)$ is the time-dependent electric field. The position operator is ill-defined in the extended Bloch states. Following the prescription of Blount~[\onlinecite{blount_1962}], the position operator in the Bloch state basis of $\hat{H}_0$, $\left\vert \psi_{n{\bm k}} \right\rangle$ is expressed as
\be
\left\langle \psi_{m{\bm k}} \right\vert \hat {r}_a \left\vert \psi_{p{\bm k}'} \right\rangle
= \left[ - \delta_{mp} i\partial_a  + \left\langle u_{m{\bm k}} \right\vert i\partial_a \left\vert u_{p{\bm k}'} \right\rangle \right] \delta({\bm k} -{\bm k}') ~.
\ee
Here, $\partial_a \equiv \partial/\partial k_a$ and $|u_{n{\bm k}}\rangle $ is the cell periodic part of the Bloch state. The second term in parentheses is the $a$ component of the interband Berry connection $\mathcal{R}_{mp}^a ({\bm k}) \equiv \langle u_{m{\bm k}} \vert i\partial_a \vert u_{p{\bm k}} \rangle$, which represents the dipole matrix elements~[\onlinecite{Sakurai2017}]. The main difficulty in using the length gauge is to deal with the highly singular function in the first term in the parenthesis. Fortunately, we can escape this term when the position operator comes as a commutator~\cite{Aversa1995}. Following the prescription of Sipe and coworkers~\cite{Aversa1995}, we can write the commutator of the position operator with any simple operator $S$ in the following form,
\be \label{iden_2}
[\bm{\hat r}, S]_{mp}=i \partial_{\bm k}S_{mp} + [\mathcal{R}_{\bm k },S]_{mp}~.
\ee
Throughout the manuscript, we use this identity extensively. For instance, the density matrix appears in the commutator form in the quantum Liouville equation. For that we use $-i[\bm{\hat r}, \rho]_{mp}=[\mathcal{D}_{\bm k}\rho]_{mp} =\partial_{\bm k}\rho_{mp} - i[\mathcal{R}_{\bm k },\rho]_{mp}$. Equipped with this tool, we now calculate the density matrix in powers of the electric field. 
%

\subsection{Perturbative solutions of the density matrix}
 
We use the quantum Liouville equation to calculate the nonequilibrium density matrix. The time evolution of the density matrix is given by 
\be \label{QLE} 
\dfrac{\partial \rho(\bm k, t)}{\partial t} + \frac{i}{\hbar}[{\cal H}, \rho(\bm k, t)] = 0 ~.
\ee 
To solve this equation, we use the adiabatic switching-on of the perturbing fields, ${\bm E}(t) = {\bm E}e^{i(\omega+i\eta) t}$. This approach, combined with the interaction picture, has been extensively used in literature for the perturbative solution of the density matrix~\cite{Boyd2008,Das2023_intrinsic}. In this work, we incorporate such adiabatic switching-on approximation as a relaxation term in the kinetic equation (See Sec.~S1 of Supplementary Material (SM)~\footnote{
\href{https://www.dropbox.com/scl/fi/vg9tn5n9cb96dnkc11hcc/SM.pdf?rlkey=56c4kwpa4zz7l2dzafkx0bccs&dl=0}
{The Supplemental material} discusses the following: i) Impact of adiabatic switching on of perturbation on the density matrix, ii) Calculation of the third-order density matrix, iii) Calculation of third-order currents for two-band model, iv) Power dissipation and Hall conductivity, v) Hermiticity and gauge invariance of the band geometric quantities, vi) Unifying extrinsic third-order quantum kinetics with semiclassics, vii) Unifying the intrinsic third-order quantum kinetics with semiclassics, and viii) Effect of same scattering time in the density matrix approach in third-order transport.} for details). For the $N$-th order $\rho$ (in powers of the electric field), Eq.~\eqref{QLE} with $\eta=1/\tau$ can be rewritten as,
\be \label{recursive_dm}
\dfrac{\partial \rho^{(N)}}{\partial t} + \dfrac{i}{\hbar}[\hat{H}_0, \rho^{(N)}] + \dfrac{\rho^{(N)}}{\tau/N} 
= \dfrac{-ie{\bm E}}{\hbar}\cdot[\bm{\hat r}, \rho^{(N-1)}] ~.
\ee
We use this equation along with Eq.~\eqref{iden_2} to obtain the density matrix of various orders in an electric field.

A few remarks regarding the third term of Eq.~\eqref{recursive_dm} are in order here. This relaxation term is known to play a crucial role in determining the shape of the resonance for the optical responses~\cite{mikhailov_PRB2015_quantum,cheng_NJP2014_third,cheng_PRB2015_third,mikhailov_PRB2016_quan,cheng_APLP2019_intra}. Although various choices of the relaxation parameters are considered in the context of {third harmonics}~\cite{nardeep_PRB2013_third, mikhailov_PRB2015_quantum, mikhailov_PRB2016_quan}, here we consider band-independent constant relaxation time that decreases by a factor $1/N$ as we go higher order in electric field. This can be understood by identifying the $E^N$ terms in the density matrix with the $N$-photon process. Physically, a $N$-photon process involves $N$ transitions that sample $N$ times energy uncertainty or level broadening for each transition. Since the scattering timescale is inverse of the energy broadening, the scattering timescale $\tau \to \tau/N$, for the density matrix contribution $\propto E^N$.  
Such a form of the relaxation term has recently proven effective for studying second-order nonlinear responses~\cite{oiwa_JPSJ2022_sys,kaplan_SP2023_uni}. It generates an intrinsic Fermi surface contribution to the Bulk photovoltaic effect~\cite{gao_PRR2021_intrin}. Furthermore, it rules out unphysical in-gap dissipative conductivity in the second-order transport~\cite{Das2023_intrinsic}.

Using Eq.~\eqref{recursive_dm} combined with Eq.~\eqref{iden_2}, we solve for the density matrix up to any arbitrary order in the electric field. The general expression of the density matrix is given by
\be \label{recursive_nth_order}
\rho^{(N)}_{mp} = \frac{e}{\hbar} g_{mp}^{-N\omega} {\bm E} \cdot [\mathcal{D}_{\bm k}\rho^{(N-1)}]_{mp} ~.
\ee 
Here, $g_{mp}^{-N\omega}=[N/\tau - i(-N\omega - \omega_{mp})]^{-1}$ inherits the scattering time dependence and $m,p$ are the band indices.
It is evident from Eq.~\eqref{recursive_nth_order} that as one goes higher order in the electric field, the expressions of the higher-order density matrix get complicated. Within the first order in an electric field, we calculate the diagonal component of the density matrix $\rho^{(1)}_{mm} = \frac{eE_d}{\hbar} g_{mm}^{-\omega} \partial_{d}f_{m} $ and off-diagonal component of the density matrix $\rho^{(1)}_{mp} = \frac{ie}{\hbar}E_d g_{mp}^{-\omega} \mathcal{R}^d_{mp}F_{mp} $. Here, $f_m$ is the equilibrium Fermi Dirac distribution function, and we have defined the difference in occupation $F_{mp}=f_m - f_p $. Following Eq.~\eqref{recursive_nth_order}, we calculate the third-order density matrix. It is given by, 
\bea \nn
\rho_{mp}^{(3)}&=& \frac{e}{\hbar} g_{mp}^{-3\omega} E_b \Big[ \mathcal{D}_{mp}^b\rho^{(2)}_{mp}
- i\mathcal{R}^b_{mp} \left(\rho^{(2)}_{pp} - \rho^{(2)}_{mm}\right) 
\\
&-& i \sum_{n\neq (m,p)} \left( \mathcal{R}^b_{mn}\rho^{(2)}_{np} -\rho^{(2)}_{mn}\mathcal{R}^b_{np}\right) \Big].
\eea
Here, $\mathcal{D}_{mp}^b=\partial_b - i(\mathcal{R}_{mm}^b - \mathcal{R}_{pp}^b)$ is the covariant derivative~[\onlinecite{Aversa1995,Ventura2017}], $\rho_{mp}^{(2)}$ and $\rho_{mm}^{(2)}$ are the interband and intraband second-order density matrices, respectively. The explicit form of the second-order density matrix is summarized in Sec.~S2 of the SM~\cite{Note1}.

\begin{figure}[t!]
    \includegraphics[width=\linewidth]{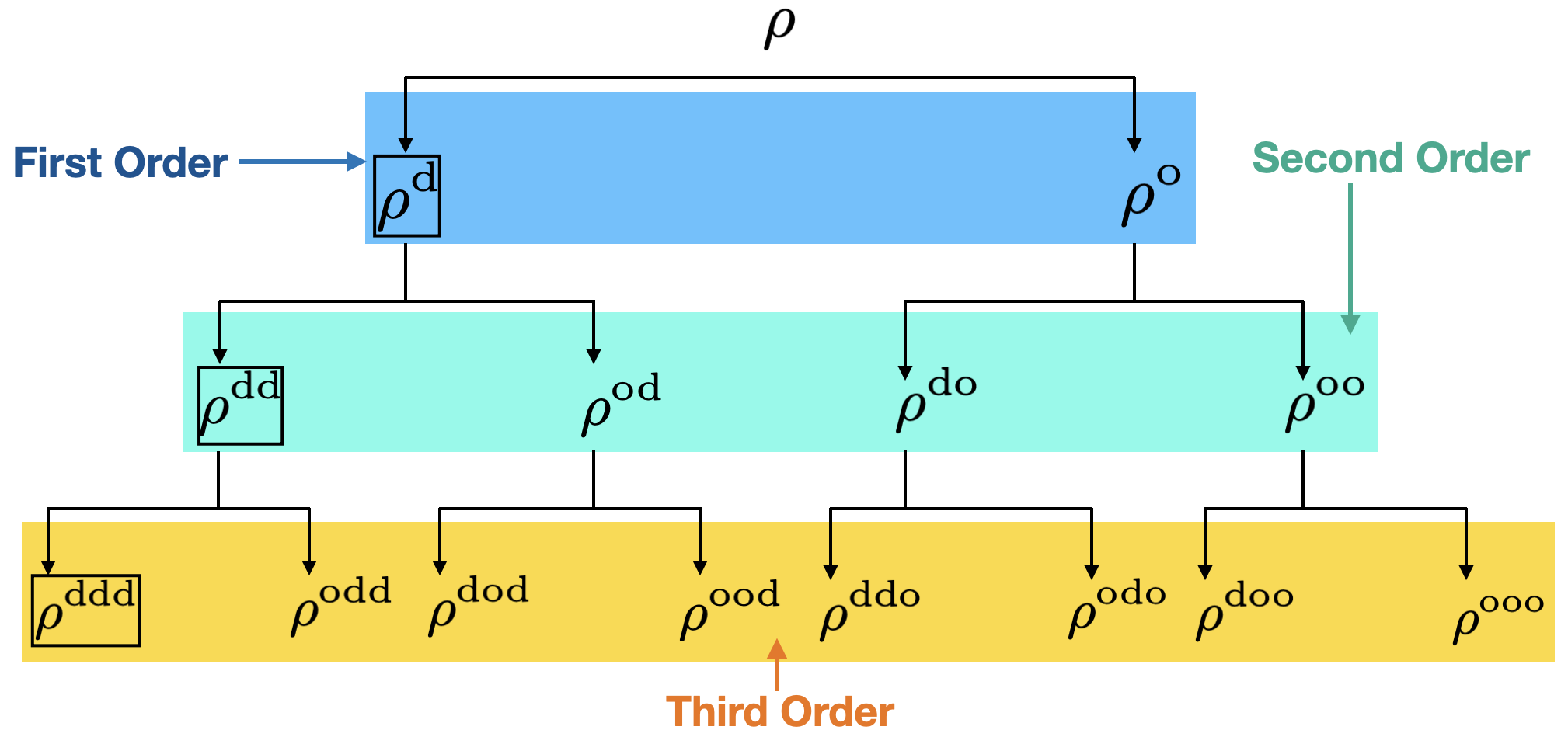}
    \caption{Tree level diagram of density matrices up to third-order in the external electric field. The diagonal (d) terms capture the modification in the distribution function. In contrast, the 
    off-diagonal (o) terms capture the impact of inter-band coherence, which manifests as a band geometric quantity. Each density matrix element contributes to the diagonal and off-diagonal components of the higher-order density matrix. The contributions in the box are the only density matrix contributions that do not involve any band geometric quantity. In contrast,  all other density matrix contributions depend on the band geometric quantities. 
    \label{fig4}}
\end{figure}

Using the expression of the second-order density matrix, we calculate the various components of the third-order density matrix as indicated in Fig.~\ref{fig4}. The detailed calculation and the expressions are presented in Sec.~S2 of the SM~\cite{Note1}. We denote the diagonal and off-diagonal elements of the linear density matrix as $\rho^{\rm d}$ and $\rho^{\rm o}$, respectively. The second order density matrix has four components, the diagonal components $\rho^{\rm dd}$, $\rho^{\rm do}$ and off-diagonal components $\rho^{\rm od}$, $\rho^{\rm oo}$. The first superscript represents the nature of the second-order density matrix component, while the second superscript denotes the first-order density matrix from which the former stems. Following similar terminology, we separate the third-order density matrix into $8$ terms. For ease of reference, we explain one of them. The density matrix $\rho_{mp}^{\rm ood}$ denotes the off-diagonal element of the third-order density matrix, which traces back to the off-diagonal second-order density matrix $\rho^{\rm od}$ and diagonal first-order density matrix $\rho^{\rm d}$.  Equipped with the expressions of all the components of the density matrix, we now calculate the third-order conductivities.

\section{Third order nonlinear conductivity}
\label{third_order_cond}
We calculate the nonlinear electric current  using the expression, 
\be \label{curr_def}
{\bm j}= -e {\rm Tr}[{\rho}\hat{\bm v}] ~.
\ee
Here, $\rho$ is the non-equilibrium density matrix which captures the impact of the electric field, and $\hat{\bm v}=\frac{1}{i\hbar}[\hat{\bm r}, \hat{H}_0]$ is the velocity operator expressed in the eigenbasis of the unperturbed Hamiltonian $\hat H_0$. The velocity operator in the eigenbasis of $\hat H_0$, is given by  $v_{pm}^a=v_m^a \delta_{pm} + (1-\delta_{pm}) i\omega_{pm} \mathcal{R}_{pm}^a$.
The diagonal part represents the band gradient velocity $v_m^a= (1/\hbar)\partial\epsilon_m/\partial k_a$, where $\epsilon_m$ is the energy of the $m$-th band. The energy difference between the $m$-th and $p$-th band is denoted by $\hbar\omega_{mp}=\epsilon_m-\epsilon_p$. The off-diagonal term signifies the interband velocity matrix, manifesting through a nonzero Berry connection $\mathcal{R}_{pm}^a$. This is finite only for multiband systems. We highlight that even though this part of the velocity is gauge-dependent, it combines with the density matrix to generate gauge-invariant physical current. 

Using Eq.~\eqref{curr_def}, the third-order current can be expressed as
\bea \label{curr_3rd}
j_a^{(3)} = - e\sum_m v_m^a \rho_{mm}^{(3)} - ie\sum_{m,p} \omega_{pm}\mathcal{R}_{pm}^a \rho_{mp}^{(3)} ~.
\eea
Here, the first term on the right-hand side represents the dispersive velocity contribution to the current, while the second term captures the Berry connection velocity contribution. 
To further simplify the expression, as shown in Fig.~\ref{fig4}, the third-order currents can be written as
\bea \label{curr_3rd_2}
j_a^{(3)} &=& -e \sum_m v_m^a \left( \rho_{mm}^{\rm ddd} + \rho_{mm}^{\rm ddo} + \rho_{mm}^{\rm dod} + \rho_{mm}^{\rm doo} \right) ~
\\
&-&  ie\sum_{m,p} \omega_{pm} \mathcal{R}_{pm}^a \left( \rho_{mp}^{\rm odd} + \rho_{mp}^{\rm odo} + \rho_{mp}^{\rm ood} + \rho_{mp}^{\rm ooo} \right)~.\nn 
\eea
We have presented detailed calculations and the expressions of the currents in Sec.~S3 of the SM~\cite{Note1}. From the expression of the current, we can extract the form of the third-order conductivity using the relation, 
\be \label{curr_3rd_def}
j_a^{(3)} = \sigma_{a;bcd} E_b E_c E_d ~.
\ee
Here, $a,b,c,d$ denote the Cartesian indices. We emphasize here that, unlike the linear conductivities, the nonlinear conductivities are not unique. There is a gauge freedom in the conductivities as an addition of terms like $\Delta \sigma_{a;bcd}=-\Delta \sigma_{a;cbd}$ keeps the current unchanged. In our paper, we take care of this gauge freedom by considering the symmetric gauge where the conductivity is symmetric in the field indices and denoted by $\bar \sigma_{a;bcd}$. From Eq.~\eqref{curr_3rd_def} it is evident that the third-order responses are reciprocal. It means that the direction of the current changes as we change the direction of the field. This is unlike the nonreciprocal second-order nonlinear response, which does not change sign upon changing the direction of the field. Importantly, the third-order response is finite even in presence of inversion symmetry. Under space inversion symmetry, the charge current and the electric field transform as ${\bm j}\to-{\bm j}$ and ${\bm E}\to -{\bm E}$, respectively. Hence, inversion symmetry breaking is not required to have a finite third-order response tensor. This contrasts the well-explored second-order nonlinear response that vanishes in inversion symmetric systems.

The limit $\omega \gg \tau$, represents the high-frequency optical response which has been explored earlier~\cite{Aversa1995,Passos2018,Parker2019,YGao2023}. Our paper considers the transport limit defined by $\omega \ll \tau$. To determine the various scattering time dependencies of the DC response, we consider the limit of $\omega\to 0$ in the functions $g_{mp}^{-N\omega}$. Then, we consider the $1/\tau \to 0$ limit and separate terms in different powers of the scattering time. To do this, we use the following expansion,
\be \label{expansion}
g_{mp}^{-N\omega} \simeq \dfrac{-i}{\omega_{mp}} + \dfrac{N}{\tau\omega_{mp}^2} + \dfrac{iN^2}{\tau^2\omega_{mp}^3} + \mathcal{O}\left( \dfrac{1}{\tau} \right)^3  ~.
\ee
Eq.~\eqref{expansion} allows us to extract all the scattering time dependence. Separating the different powers of the scattering time dependences, the third-order transport current can be expanded as $j^{(3)}=j^{(3)}(\tau^0)+j^{(3)}(\tau)+j^{(3)}(\tau^2)+j^{(3)}(\tau^3)$. See Table.~S1 in  Sec.~S3 of the SM~\cite{Note1} for details. Interestingly, the time-reversal symmetry determines which order (even or odd) of the scattering time contributes to a nonzero current. This can be understood as follows. Under time reversal symmetry, ${\bm j} \to -{\bm j}$, ${\bm E}\to {\bm E}$ and $\tau\to -\tau$. Consequently, contributions with odd powers of $\tau$ survive in time-reversal symmetric systems, while the current contributions with even powers of $\tau$ vanish.

In addition to the separation in scattering time dependence, we segregate the third-order conductivities into Ohmic or dissipative and non-ohmic or dissipationless parts. The latter one is determined by the `no work done' criteria, or ${\bm j}^{(3)} \cdot {\bm E} = 0$ (see Sec.~S4 of the SM~\cite{Note1}). This condition is easily satisfied by the conductivities that are anti-symmetric on exchanging the field and current indices. A straightforward implication of this is that the purely longitudinal conductivity $(a=b=c=d)$ is always Ohmic or dissipative. We use the symmetry or antisymmetry property of the current and field indices in the conductivity to separate conductivities into dissipationless and dissipative parts. A formal and equivalent way to segregate the dissipative and non-dissipative part of the conductivity has been recently proposed by Tsirkin and Souza~\cite{Tsirkin2022}. For a general expression of current $j_{a_1}^{(N)}=\sigma_{a_1;a_2..,a_{n+1}}^{(N)}E_{a_2}E_{a_3}..E_{a_{n+1}}$, it has been shown that the dissipative part of the $N$-th order conductivity can be obtained by symmetrizing over all the spatial indices~
\be \label{def_dis}
\sigma_{a_1;a_2..,a_{n+1}}^{N\rm O} =\dfrac{1}{(n+1)!}\sum_{a_1,..,a_{n+1}} \sigma_{a_1,..,a_{n+1}}^{(N)}~.
\ee
The non-dissipative conductivity is then obtained as 
\be \label{def_Hall}
\sigma_{a_1;a_2..,a_{n+1}}^{N\rm H}= \sigma_{a_1;a_2..,a_{n+1}}^{(N)}-\sigma_{a_1;a_2..,a_{n+1}}^{N\rm O} ~.
\ee
Here, $\sigma^{N\rm O}$ has been identified as the Ohmic conductivity, and $\sigma^{N\rm H}$ as the Hall conductivity. In the following, we will discuss third-order conductivity only; hence, the superscript ``$N$" will be omitted.

Interestingly, we can express all the third-order conductivities in terms of three different band geometric quantities~\cite{Provost1980}. Namely, the band-resolved quantum metric - $\mathcal{G}_{mp}^{ab}$, the band-resolved Berry curvature (BC) - $\Omega_{mp}^{ab}$, and the symplectic connection (SC) - $\tilde{\Gamma}_{mp}^{abc}$. All of these quantities are imaginary (or real) parts of more general quantities, the quantum geometric tensor - $\mathcal{Q}_{mp}^{ab}$ ~\cite{Bhalla2022, Fregoso2019}, and the geometric connection - $\mathcal{C}_{mp}^{abc}$ \cite{Ahn2020,Watanabe2021} defined as
\bea
\mathcal{Q}_{mp}^{ab} &=& \mathcal{R}_{pm}^a \mathcal{R}_{mp}^b = \mathcal{G}_{mp}^{ab} - \frac{i}{2} \Omega_{mp}^{ab}~,
\\
\mathcal{C}_{mp}^{abc} &=& \mathcal{R}_{pm}^a \mathcal{D}_{mp}^b\mathcal{R}_{mp}^c = \Gamma_{mp}^{abc} - i\tilde{\Gamma}_{mp}^{abc} ~.
\eea
For more details and pedagogy of band geometry of the Bloch electrons, see Ref.~[\onlinecite{Ahn2021,Smith2022_momentum_gravity,Hetenyi2023_fluctuation}]. The Hermiticity and the gauge invariance of the band geometric quantities are shown in Sec.~S5 of the SM~\cite{Note1}. Below, we first calculate the extrinsic $\tau^3$, $\tau^2$ and linear-$\tau$-dependent third order conductivities~\cite{XLiu2023_covariant, Parker2019, HLiu2022_BCP_prb}. Then, we move to our central finding related to the contributions independent of the scattering time.

\subsection{Extrinsic \texorpdfstring{$\tau^3$}{TEXT} contribution}
We find that only the $\rho^{\rm ddd}$ component of the density matrix contributes to the cubic scattering time-dependent third-order current. We extract the field symmetric conductivity as
\be\label{NL_Drude}
\bar{\sigma}_{a;bcd} (\tau^3) = -\dfrac{e^4\tau^3}{6\hbar^4} \sum_{m,p, {\bm k}}  (\partial_{a} \epsilon_m) \partial_{bcd} f_m ~.
\ee
This dissipative conductivity was previously derived as a third-order nonlinear Drude current by solving the Boltzmann equation in Ref.~\cite{XLiu2023_covariant}. This conductivity can be thought of as a result of the Fermi sea shift in the third order of the electric field. Under time reversal, the energy satisfies $\epsilon(-{\bm k})=\epsilon({\bm k})$ and since there are four-momentum derivatives, the integration above is even under momentum reversal and hence finite in the presence of time-reversal symmetry. This is consistent with our previous argument related to the odd power of scattering time.

\subsection{Extrinsic \texorpdfstring{$\tau^2$}{TEXT} contribution}
The quadratic scattering time-dependent current originates from the density matrix elements $\rho^{\rm ddo}, \rho^{\rm dod}$ and $\rho^{\rm odd}$. After performing some straightforward but lengthy algebra, we obtain the following form, 
\be\label{sigma_quad_tau}
\bar{\sigma}_{a;bcd}(\tau^2) = \dfrac{\tau^2e^4}{6\hbar^3} \sum_{m,p, {\bm k}} \left( \partial_{cd}^2 \Omega_{mp}^{ab} + \partial_{db}^2 \Omega_{mp}^{ac} + \partial_{bc}^2 \Omega_{mp}^{ad} \right) f_m .
\ee
This conductivity is Hall in nature. This can be understood from the presence of the Berry curvature tensor, an anti-symmetric tensor. The presence of a second-order derivative of the Berry curvature in the expression means that this contribution to the Hall current originates from the Berry curvature quadrupole~[\onlinecite{Zhang2023_multipole}]. Such a Berry curvature quadrupole contribution to the third-order current has been previously obtained in the asymmetrised form using semiclassical wavepacket theory~\cite{XLiu2023_covariant,Sorn2023_quadrupole}, the quantum diagrammatic approach~\cite{Parker2019} and experimentally demonstrated in kagome antiferromagnet~\cite{sankar_arxiv2023_observation}. Here, we provide the derivation of this conductivity using the density matrix approach for the first time. Under time-reversal, the Berry curvature is an odd function of momentum ${\bm \Omega}(-{\bm k})=-{\bm \Omega}({\bm k})$; hence this contribution is finite only in magnetic systems where time-reversal symmetry is broken. Recently, it has been proposed that it can probe hidden octopolar order in metallic system~\cite{Sorn2023_quadrupole}.

\subsection{Extrinsic \texorpdfstring{$\tau^1$}{TEXT} contribution}
The linear-$\tau$-dependent current originates from the density matrices $\rho^{\rm ddo}$, $\rho^{\rm dod}$, $\rho^{\rm doo}$, $\rho^{\rm odd}$, and $\rho^{\rm ood}$.
This extrinsic third-order conductivity is governed by the normalized and band-resolved quantum metric $\tilde {\mathcal G}_{mp}^{ab}=\mathcal {G}_{mp}^{ab}/\omega_{mp}$, often associated with Berry connection polarizability (BCP) tensor. We calculate the conductivity to be, 
\bea \label{sigma_qm_linear_tau}
\bar{\sigma}_{a;bcd} (\tau) &=& \tau \dfrac{e^4}{3\hbar^3} \sum_{m,p, {\bm k}} \left[\partial_{ad}^2  \tilde {\mathcal G}_{mp}^{bc}+\partial_{ac}^2  \tilde {\mathcal G}_{mp}^{bd}+\partial_{ab}^2 \tilde {\mathcal G}_{mp}^{cd} \right.\nn
\\
&-& \left. 3\left(\partial_{cd}^2 \tilde {\mathcal G}_{mp}^{ab} + \partial_{bd}^2 \tilde {\mathcal G}_{mp}^{ac} +\partial_{bc}^2 \tilde {\mathcal G}_{mp}^{ad}\right)\right]f_m~.
\eea
This is the band-normalized quantum metric quadrupole-induced third-order conductivity. Since the quantum metric is an even function under both time-reversal and inversion symmetry $\mathcal{G}_{mp}^{ab}(-{\bm k})=\mathcal{G}_{mp}^{ab}({\bm k})$, and momentum is an odd function, Eq.~\eqref{sigma_qm_linear_tau} can be finite in nonmagnetic and centrosymmetric systems. Such linear $\tau$-dependent conductivity proportional to the quantum metric quadrupole was earlier obtained in its asymmetrized form using a semiclassical wave-packet approach in Ref.~[\onlinecite{HLiu2022_BCP_prb}] and using quantum transport theory in Ref.~\cite{Wei_PRB2022_quantum}. Here, we derived it using the quantum kinetic density matrix approach. We mention that our derivation is limited to two-band systems, neglecting multiband contributions. We note that the real part of the geometric connection, $\Gamma_{mp}^{abc}$, is related to the dipole of the quantum metric $\partial_a \mathcal{G}_{mp}^{bc}=\Gamma_{mp}^{bac}+\Gamma_{mp}^{cab}$. Notably, the linear-$\tau$ contribution is of Fermi surface nature and there are no additional Fermi sea terms. The proper choice of relaxation time plays a crucial role in eliminating any unphysical in-gap dissipative conductivity. We have discussed this issue in detail in section~\ref{discuss}.

We further segregate Eq.~\eqref{sigma_qm_linear_tau} in its Ohmic (dissipative) and non-Ohmic (non-dissipative) contributions. The dissipationless (Hall) conductivity linear in scattering time is given by
\bea \label{lin_tau_hall}
\bar{\sigma}_{a;bcd}^{\rm H} (\tau) &=& \tau \dfrac{2e^4}{3\hbar^3} \sum_{m,p,{\bm k}} \left[\partial_{ad}^2 \tilde {\mathcal G}_{mp}^{bc} + \partial_{ab}^2 \tilde {\mathcal G}_{mp}^{cd} + \partial_{ac}^2 \tilde {\mathcal G}_{mp}^{bd} \right.\nn
\\
&-& \left. \partial_{cd}^2 \tilde {\mathcal G}_{mp}^{ab} - \partial_{bd}^2 \tilde {\mathcal G}_{mp}^{ac} - \partial_{bc}^2 \tilde {\mathcal G}_{mp}^{ad} \right] f_m ~.
\eea 
This Hall conductivity is extremely significant as it is the only Hall response up to the third order in an electric field, which is finite even in systems with both fundamental symmetries. We use ``$\rm H$" in the superscript to define the dissipationless or Hall part of the conductivity. Such third-order Hall current was recently experimentally observed in TMDCs~\cite{Lai2021,Ye2022_orbital}. 
The dissipative (Ohmic) conductivity linear in scattering time is given by
\bea \label{cond_lin_tau_Ohm}
\bar{\sigma}_{a;bcd}^{\rm O} (\tau) &=& -\tau \dfrac{e^4}{3\hbar^3} \sum_{m,p,{\bm k}} \left[\partial_{cd}^2 \tilde{\mathcal {G}}_{mp}^{ab} + \partial_{bd}^2 \tilde{\mathcal {G}}_{mp}^{ac} + \partial_{bc}^2 \tilde{\mathcal {G}}_{mp}^{ad} \right.\nn
\\
&+& \left. \partial_{ad}^2\tilde{\mathcal {G}}_{mp}^{bc} + \partial_{ac}^2 \tilde{\mathcal {G}}_{mp}^{bd} + \partial_{ab}^2 \tilde{\mathcal {G}}_{mp}^{cd}\right] f_m ~.
\eea
This is the quantum geometry-induced dissipative transport coefficient, which can have both longitudinal and transverse contributions. We use ``$\rm O$" in the superscript to define the dissipative or Ohmic part of the conductivity. So far, this contribution has not been experimentally realized. Equations~\eqref{lin_tau_hall} and \eqref{cond_lin_tau_Ohm} unify the quantum geometry induced nonlinear Hall and dissipative responses in inversion symmetric systems.

\subsection{Intrinsic third order conductivity\label{intrnsc_third}}

Now, we move on to the paper's main focus, the scattering time-independent contributions. Unlike the rest of the coefficients discussed above, intrinsic conductivities can be precisely calculated through ab initio methods and may be linked to material-specific topological invariants. The total third-order intrinsic contribution can be expressed as a sum of Berry curvature and a symplectic connection contribution, 
\be
\bar{\sigma}_{a;bcd} (\tau^0) = \bar{\sigma}_{a;bcd}^{\rm BC} + \bar{\sigma}_{a;bcd}^{\rm SC} ~.
\ee
The BC contribution to the conductivity is obtained to be
\be \label{sigma_bc_sym}
\bar{\sigma}_{a;bcd}^{\rm BC} = \dfrac{e^4}{\hbar^3} \sum_{m,p, {\bm k}} \dfrac{-1}{\omega_{mp}^2} (\Omega_{mp}^{ab} \partial_{cd}^2 + \Omega_{mp}^{ac} \partial_{bd}^2  + \Omega_{mp}^{ad} \partial_{bc}^2) f_m ~.
\ee
This contribution originates from the third-order density matrix term, $\rho^{\rm odd}$ and can be rewritten as a quadrupole moment of the band normalized Berry curvature $\Omega_{mp}/\omega_{mp}^2$. As far as we know, such an intrinsic contribution to the third-order response has not been explored earlier. The intrinsic conductivity from the SC is obtained to be 
\bea \label{sigma_sc_sym}
\bar{\sigma}_{a;bcd}^{\rm SC}
&=& \dfrac{e^4}{\hbar^3} \sum_{m,p, {\bm k}} \dfrac{1}{\omega_{mp}^2} 
\left[  \tilde{\Gamma}_{mp}^{\overline{abc}} \partial_d f_m
+ \tilde{\Gamma}_{mp}^{\overline{adb}} \partial_c f_m \right. \nn
\\ \label{sigma_sc_sym}
&+&  \left.  \tilde{\Gamma}_{mp}^{\overline{acd}} \partial_b f_m - \dfrac{1}{3}\tilde{\Gamma}_{mp}^{\overline{bcd}}  \partial_a f_m \right] ~.
\eea
Here, we have defined $\tilde{\Gamma}_{mp}^{\overline{abc}}$ as the $a,b,c$ symmetric SC: $\tilde{\Gamma}_{mp}^{\overline{abc}}=\tilde{\Gamma}_{mp}^{abc}+\tilde{\Gamma}_{mp}^{bca}+\tilde{\Gamma}_{mp}^{cab}$ for brevity.
The contribution to Eq.~\eqref{sigma_sc_sym} stems from the density matrices $\rho^{\rm doo}$, $\rho^{\rm ood}$ and $\rho^{\rm ooo}$. 
We emphasize here that although an intrinsic dissipationless contribution has been predicted recently in Ref.~\cite{Xiang2023}, Eq.~\eqref{sigma_sc_sym} is a new result, as will be highlighted in the next few paragraphs.  
Equation~\eqref{sigma_bc_sym}-\eqref{sigma_sc_sym} are the main results of this paper.

We mention here that for the intrinsic contributions, we restrict ourselves only to a two-band system. See Sec.~S3 of the SM~\cite{Note1} for details. Interestingly, we find that third-order responses are Fermi surface in nature and all the Fermi sea contributions from different density matrix contributions add up to zero. The different scattering time dependence of the various harmonics plays a crucial role in making the total Fermi sea contribution vanish. We discuss this in detail in Sec.~\ref{discuss}. Note that the intrinsic conductivities are finite only when the time reversal (TR) symmetry is broken. In TR preserving systems, the energy is an even function of momentum, and the BC is an odd function of momentum. At the same time, the double derivative in Eq.~\eqref{sigma_bc_sym} is an even function of momentum. Thus, $\bar{\sigma}^{\rm BC}({\bm k}) =  -\bar{\sigma}^{\rm BC}(-{\bm k})$. Consequently, the BC contribution to the intrinsic third-order response vanishes. Similarly, the SC is an even function of momentum for TR-preserving systems, and the momentum derivative is an odd function. Thus, we have $\bar{\sigma}^{\rm SC}({\bm k}) =  -\bar{\sigma}^{\rm SC}(-{\bm k})$, and the SC contribution also vanishes in TR preserving systems. This establishes that the TR symmetry must be broken to obtain a net intrinsic third-order response.

Now, we segregate the intrinsic conductivities into Ohmic and Hall parts. The conductivity given in Eq.~\eqref{sigma_bc_sym} shows the antisymmetry property originating from the property of Berry curvature. Hence, the BC-induced current given by Eq.~\eqref{sigma_bc_sym} is purely Hall (or non-dissipative). However, this is not true for the SC conductivity. Using the symmetry or antisymmetry property of the current and field indices in the conductivity, Eq.~\eqref{sigma_sc_sym} can be separated into the dissipationless and dissipative parts straightforwardly. The conductivity with antisymmetry property is separated as 
\bea \label{sigma_sc_Hall}
\bar \sigma_{a;bcd}^{\rm H,SC} &=& \dfrac{e^4}{\hbar^3} \sum_{m,p, {\bm k}} \dfrac{1}{3\omega_{mp}^2} \left[\tilde{\Gamma}_{mp}^{\overline{cda}} \partial_b  -\tilde{\Gamma}_{mp}^{\overline{bcd}}  \partial_a  
 \right. \nn
\\
&+&
\left.  \tilde{\Gamma}_{mp}^{\overline{dab}} \partial_c -\tilde{\Gamma}_{mp}^{\overline{bcd}}  \partial_a   + \tilde{\Gamma}_{mp}^{\overline{abc}} \partial_d 
- \tilde{\Gamma}_{mp}^{\overline{bcd}}  \partial_a  
\right] f_m ~.~~~~
\eea
The longitudinal counterpart of this conductivity vanishes. This dissipationless conductivity has been recently derived using the semiclassical wave-packet approach in Ref.~\cite{Xiang2023}. Here, we reproduce their result using the quantum kinetic theory framework for the first time. Therefore, the total intrinsic Hall conductivity can be obtained from  \eqref{sigma_bc_sym} and \eqref{sigma_sc_Hall} and formally can be written as $\bar \sigma_{a;bcd}^{{\rm H}} = \bar \sigma_{a;bcd}^{\rm BC} + \bar \sigma_{a;bcd}^{\rm H,SC}$. The dissipative part of Eq.~\eqref{sigma_sc_sym} is given by
\bea \label{curr_intrnsc_long}
\bar \sigma_{a;bcd}^{\rm O,SC}
&=& \dfrac{e^4}{\hbar^3} \sum_{m,p, {\bm k}} \dfrac{2}{3\omega_{mp}^2} \left[ \tilde{\Gamma}_{mp}^{\overline{bcd}}  \partial_a
+ \tilde{\Gamma}_{mp}^{\overline{cda}} \partial_b  \right. \nn
\\ 
&+& \left. \tilde{\Gamma}_{mp}^{\overline{dab}} \partial_c  +\tilde{\Gamma}_{mp}^{\overline{abc}} \partial_d \right] f_m .
\eea
This is one of the main results of this paper. This contribution to the dissipative Fermi surface conductivity, to the best of our knowledge, is predicted here for the first time. The key thing to note here is that the conductivity in Eq.~\eqref{curr_intrnsc_long} is completely cyclic in all the current and the field indices. This assures that this is a purely dissipative response. Combined Eq.~\eqref{sigma_sc_Hall} and \eqref{curr_intrnsc_long} imply a merger of the reciprocal nonlinear Hall and magnetoresistance induced by quantum geometry in inversion symmetric systems.

\section{Unification with semiclassical wave-packet approach \label{unifying_semiclasscs}}

In this section, we reconcile the third-order conductivities obtained from our quantum kinetic theory with the semiclassical wave-packet approach results~\cite{HLiu2022_BCP_prb,Zhang2023_multipole,Xiang2023,Nag2022,Wei_PRB2022_quantum,YGao2023,Fang2023_altermagnet}. For the sake of completeness, we will briefly sketch the semiclassical wave-packet approach. For detailed intricacies of the semiclassical wave-packet method, we direct readers to excellent Refs.~\cite{xiao_RMP2010_berry,gao_FP2019_semiclass,Xiang2023}. In semiclassics, the transport current is calculated using the combination of semiclassical wave-packet dynamics and the Boltzmann transport equation.  The semiclassical equations of motion, accurate
up to third-order, are given by~\cite{xiao_RMP2010_berry,gao_FP2019_semiclass,Xiang2023} (we set $e=\hbar=1$)
\bea
{\bm{\dot r}} = \frac{\partial \tilde {\epsilon}}{\partial {\bm k}} -  {\bm {\dot{k}}} \times {\bm {\tilde{ \Omega}}}; ~~~~ {\bm {\dot k}}= - {\bm E}~.
\eea
Here, $\tilde{\epsilon}$ and $\tilde{\bm\Omega}$ are the total energy and Berry curvature of a single band, incorporating the electric field-induced corrections. Since ${\bm {\dot k}}$ explicitly contains ${\bm E}$ in its expression, the Berry curvature corrected up to second-order 
\be
{\bm {\tilde{\Omega}}}={\bm \Omega}+{\bm \Omega}^{(1)}+{\bm \Omega}^{(2)},
\ee
makes $ {\bm{\dot r}}$ accurate to third-order in electric field. Here, ${\bm \Omega}$ is the conventional Berry curvature and ${\bm \Omega}^{(1/2)}$ represents the first-/second-order field correction to it. The correction to the Berry curvature can be traced back to the field-induced positional shift of the center of mass of the wave packet within the unit cell. Up to the second order in an electric field, this is given by
\be \label{connection}
{\bm r}_c = \frac{\partial \gamma}{\partial {\bm k}} + {\bm {\mathcal A}} +  {\bm {\mathcal A}}^{(1)} + {\bm {\mathcal A}}^{(2)}~.
\ee
Here, $\gamma$ is the phase factor of the wave-packets envelop function, ${\bm {\mathcal A}}$ is the usual Berry connection, ${\bm {\mathcal A}}^{(1)}$ is the first-order positional shift introduced by Gao in Ref.~[\onlinecite{YGao2014}] and ${\bm {\mathcal A}}^{(2)}$ is the second-order positional shift introduced in Ref.~\cite{Xiang2023}. By taking the curl of ${\bm {\mathcal A}}$, ${\bm {\mathcal A}}^{(1)}$, and ${\bm {\mathcal A}}^{(2)}$, we get  ${\bm \Omega}={\bm \nabla}_{\bm k}\times {\bm {\mathcal A}}$, ${\bm \Omega}^{(1)}={\bm \nabla}_{\bm k}\times {\bm {\mathcal A}}^{(1)}$, and ${\bm \Omega}^{(2)}={\bm \nabla}_{\bm k}\times {\bm {\mathcal A}}^{(2)}$, respectively. In addition to the correction to the anomalous velocity, the quasiparticle energy $\tilde{\epsilon}$ also has electric field correction. Accurate up to third-order in ${\bm E}$ it is given by~[\onlinecite{YGao2014,HLiu2022_BCP_prb,Xiang2023}],
\bea 
\tilde{\epsilon}=\epsilon+\epsilon^{(2)}+\epsilon^{(3)}~,
\eea 
where $\epsilon$ is the unperturbed band energy and $\epsilon^{(2/3)}$ are the second-/third-order field correction to energy. Note that the energy does not have a linear in-field correction. From the perturbation theory, the first-order energy correction $\langle m \lvert e{\bm E}\cdot({\bm r}-{\bm r}_c)\rvert m\rangle$ can be shown to vanish~[\onlinecite{Xiang2023}], where ${\bm r}_c$ is the position of the center of the wave packet.

In addition to the semiclassical dynamics, the non-equilibrium distribution function is needed to calculate the current. Within the relaxation time approximation~\cite{ashcroft_book1976_solid}, the nonequilibrium distribution function $g({\bm k})$ in the steady state is solved from,
\be
\dot{\bm k}\cdot\nabla_{\bm k}g=-\frac{g-f}{\tau}~,
\ee
where $f$ is the equilibrium Fermi-Dirac distribution function and $\tau$ is the scattering time. Combined with the semiclassical equations of motion, the distribution function is obtained as a series expansion
\bea 
g=f(\epsilon) + \sum_{\alpha=1}^{\infty}(\tau{\bm E}\cdot\nabla_{\bm k})^\alpha f(\epsilon)~.
\eea
{Note that in the above equation, we consider equilibrium energy $\epsilon$, not the electric field modified energy $\tilde \epsilon$, in the Fermi-Dirac distribution function. This differs from the previous literature~\cite{HLiu2022_BCP_prb,Nag2022,Wei_PRB2022_quantum}.
We discuss the reasons for this choice in detail in section~\ref{discuss}.}

Equipped with this, we now calculate the different orders of $\tau$-dependent current. The nonlinear Drude contribution arises from the third-order modification in the distribution function and is expressed as~\cite{HLiu2022_BCP_prb,XLiu2023_covariant}
\be \label{NL_Drude_SC}
\bar{\sigma}_{a;bcd}(\tau^3)=-\tau^3\sum_{m,{\bm k}} (\partial_a\epsilon_m)\partial^3_{bcd} f_m~.
\ee
This expression aligns with Eq.~\eqref{NL_Drude} apart from the factor related to the relaxation time. The anomalous velocity first order in the electric field gives rise to the third order current proportional to $\tau^2$. In the field-symmetrized form, it is given by~\cite{XLiu2023_covariant,Fang2023_altermagnet},
\be  \label{cond_quad_tau_BC}
\bar{\sigma}_{a;bcd} (\tau^2) = \dfrac{\tau^2}{3} \sum_{m,{\bm k}} f_m \left[ \partial_{bc}^2 \Omega_m^{ad} + \partial_{cd}^2 \Omega_m^{ab} + \partial_{db}^2 \Omega_m^{ac} \right] ~.
\ee
This expression broadly matches with Eq.~\eqref{sigma_quad_tau} except for a factor related to the choice of relaxation time for the calculation of nonlinear responses (see Sec.~S6 of SM~\cite{Note1}, and Sec.~S1 for a discussion on the choice of the scattering timescale). 

The third-order current linear in $\tau$ arises from the first-order field correction to the BC and the second-order energy correction in the dispersive velocity. In Eq.~\eqref{connection}, the field correction to the Berry connection up to the first order is given by,
\bea
{\mathcal A}_m^{a, (1)}({\bm k})=G_m^{ab} ({\bm k})E_b~,
\eea 
where $G_m^{ab}=2\sum_{p\neq m} \mathcal{G}_{mp}^{ab}/\omega_{mp}$ is the Berry connection polarizability (BCP) tensor. The resultant anomalous velocity $ {\bm E} \times {\bm \Omega}^{(1)}$ is second order in electric field. Together with the first-order non-equilibrium distribution function, the first-order correction to Berry curvature gives rise to the following third-order conductivity~[\onlinecite{HLiu2022_BCP_prb,Fang2023_altermagnet}],
\bea \label{cond_lin_tau_BC1}
\bar{\sigma}_{a;bcd}^{\rm An} (\tau) &=& \dfrac{\tau}{3} \sum_{m,{\bm k}} f_m \left[(\partial_{ad}^2 G_m^{bc} + \partial_{ab}^2 G_m^{cd} + \partial_{ac}^2 G_m^{bd}) \right.\nn
\\
&-& \left.(\partial_{cd}^2 G_m^{ab} + \partial_{bd}^2 G_m^{ac} + \partial_{bc}^2 G_m^{ad}) \right] ~.
\eea 
This current is completely Hall-like and identical to the Hall current in Eq.~\eqref{lin_tau_hall} obtained from quantum kinetic theory (Sec.~S6 of SM~\cite{Note1}). Another contribution to the linear in $\tau$ third-order current arises from the second-order energy correction~[\onlinecite{YGao2014,HLiu2022_BCP_prb}],
\bea 
\epsilon_m^{(2)}= - \frac{1}{2} E_a G_m^{ab} E_b~.
\eea 
This energy correction modifies the dispersive velocity that has been recently shown to induce a second-order nonlinear conductivity~\cite{Kaplan2023_unification}.  The first-order non-equilibrium distribution function then leads to the third-order conductivity given by,
\be \label{cond_lin_tau_enrgy2}
\bar{\sigma}_{a;bcd}^{\bar{\epsilon}}(\tau) = - \frac{\tau}{6} \sum_{m, {\bm k}} f_m \left[ \partial^2_{ab}G_m^{cd} + \partial^2_{ad}G_m^{bc} + \partial^2_{ac}G_m^{bd} \right] ~.~~ 
\ee 
Unlike the BC correction-induced conductivity, Eq.~\eqref{cond_lin_tau_enrgy2} is neither dissipative nor dissipationless. The dissipative (symmetric) part of this conductivity is separated as,
\bea \label{cond_lin_tau_Ohm_enrgy2}
\bar{\sigma}^{\rm O,\bar{\epsilon}}_{a;bcd}(\tau) &=& -\frac{\tau}{12} \sum_{m ,{\bm k}} f_m \left[ \partial_{ab}^2 G_m^{cd} + \partial_{ac}^2 G_m^{bd} + \partial_{ad}^2 G_m^{bc}  \right. \nn
\\
&+& \partial_{bc}^2 G_m^{ad} + \left. \partial_{bd}^2 G_m^{ac} + \partial_{cd}^2 G_m^{ab} \right] ~.
\eea 
The above expression matches with the linear $\tau$ third-order conductivity in Eq.~\eqref{cond_lin_tau_Ohm} obtained from the quantum kinetic theory, except for a factor from relaxation time.  
The Hall (antisymmetric) part from Eq.~\eqref{cond_lin_tau_enrgy2} is given by,
\bea \label{cond_lin_tau_Hall_enrgy2}
\bar{\sigma}^{\rm H,\bar{\epsilon}}_{a;bcd}(\tau) &=& \frac{\tau}{12} \sum_{m, {\bm k}} f_m \left[ \partial_{bc}^2 G_m^{ad} - \partial_{ac}^2 G_m^{bd} + \partial_{cd}^2 G_m^{ab} \right. \nn
\\
&-& \partial_{ad}^2 G_m^{bc}+ \left. \partial_{db}^2 G_m^{ac} - \partial_{ab}^2 G_m^{cd} \right] ~.
\eea
We find that the contribution in the above expression does not have an analog in our quantum kinetic theory.

Finally, we will compare the intrinsic third-order conductivities obtained from these two formalisms. The second-order positional shift can be expressed in terms of the second-order Berry connection polarizability $T_{a b c} $ as ${\mathcal A}^{a,(2)}=T_{a b c} E_b E_c$, where 
\be 
T^m_{a b c} ={\rm Re} \sum_{p \neq m} \bigg( {\mathcal U}_{mp}^{a b c} + {\mathcal U}_{mp}^{ b a c}-  {\mathcal U}_{mp}^{ b c a } - \sum_{n \neq (m,p)} {\mathcal V}_{mpn}^{a b c} \bigg)~.
\ee
Here, the superscript $m$ is the band index of the band in which we are interested, and the sum on $p$ represents all other bands. The two-band quantity ${\mathcal U}_{mp}^{a b c}$ and multiband quantity ${\mathcal V}_{mpn}^{a b c}$ are specified by 
\bea
{\mathcal U}_{mp}^{a b c}&=&M_{mp}^{b} \left({\mathcal A}_m - {\mathcal A}_p - i \partial_a \right) M_{pm}^{c}~,
\\
{\mathcal V}_{mpn}^{a b c}&=& \left(2 M_{mp}^{a} {\mathcal A}_{pn}^{b} + M_{mp}^{b} {\mathcal A}_{pn}^{a} \right)  M_{nm}^{c}~.
\eea
Here, $M_{mp}^a=\mathcal{A}_{mp}^a/(\epsilon_p - \epsilon_m)\Bar{\delta}_{pm}$ with $\Bar{\delta}_{pm}=1-\delta_{pm}$. The third order correction to the velocity $ {\bm E} \times {\bm \Omega}^{(2)}$, gives rise to the following Hall conductivity [\onlinecite{Xiang2023}],
\be
\sigma_{a;bcd} (\tau^0) = \sum_{m, {\bm k}} \left( \partial_b T^m_{acd} - \partial_a T^m_{bcd} \right) f_m ~.
\ee
To unify with our results, we first symmetrize the conductivity in the field indices. Following this, we simplify the results in terms of the band geometric quantities introduced in this paper. Note that the two-band quantity ${\mathcal U}_{mp}^{a b c}$ can be simplified in terms of quantum geometric tensor and geometric connection, as shown in Sec.~S7 of SM~\cite{Note1}. Ignoring the multiband term ${\mathcal V}_{mpn}^{a b c}$, the third-order intrinsic Hall conductivity $\sigma_{a;bcd}$ in the symmetric gauge can be written as 
\bea \label{curr_intrinsic_Xiang23} \nn
\bar \sigma_{a;bcd} (\tau^0) &=& \sum_{m,p, {\bm k}} \dfrac{1}{3\omega_{mp}^2} \left[ \tilde{\Gamma}^{\overline{cad}}_{mp}  \partial_b 
+ \tilde{\Gamma}^{\overline{bad}}_{mp} \partial_c + \tilde{\Gamma}^{\overline{cab}}_{mp} \partial_d 
\right. ~\nn
\\
&-& \left. 3\tilde{\Gamma}^{\overline{cbd}}_{mp} \partial_a \right] f_m ~.
\eea
To our satisfaction, this Hall conductivity is identical to the contribution from the symplectic connection of the total Hall conductivity derived from our quantum kinetic theory in Eq.~\eqref{sigma_sc_Hall}. The gauge-invariant energy correction in the third order of the electric field is obtained to be ~\cite{Xiang2023}
\bea
\bar{\epsilon}_m^{(3)} &=& {\rm Re}\sum_p\left[\left(\mathcal{U}_{pm}^{bca} - \mathcal{U}_{mp}^{bac}\right) + \sum_{ n \neq p}  2 M_{mp}^a\mathcal{A}_{pn}^b M_{nm}^c \right] ~\nn
\\
&& E_a E_b E_c ~.
\eea
Here, the first term is a two-band contribution, and the second is a multiband term contribution. Neglecting the multiband contribution, we can express the energy correction in terms of the symplectic connection as $\Bar{\epsilon}^{(3)}_m = -\sum_p \frac{1}{\omega_{pm}^2} \left( \tilde{\Gamma}^{cba}_{mp} +\tilde{\Gamma}^{abc}_{mp}\right) E_a E_b E_c$. This contribution has not been explored to date. We find that the field-symmetrized intrinsic third-order conductivity originating from the energy correction is given by
\be \label{energy_curr_third}
\bar{\sigma}_{a;bcd}^{\bar \epsilon} (\tau^0) = \frac{2}{3}\sum_{p, m, {\bm k}} \frac{1}{\omega_{mp}^2}\left(\tilde{\Gamma}^{bcd}_{mp} +\tilde{\Gamma}^{cdb}_{mp}+\tilde{\Gamma}^{dbc}_{mp}\right)\partial_a f_m~.
\ee
Note that this contribution can neither be classified as Hall nor as dissipative. So, we separate these contributions using Tsirkin and Souza's method~\cite{Tsirkin2022}. The dissipative conductivity is obtained to be 
\bea \label{energy_dissipative}
\bar \sigma_{a;bcd}^{{\rm O},\bar{\epsilon}} (\tau^0) &=& \sum_{p, m, {\bm k}} \dfrac{1}{6\omega_{mp}^2} \left[ \tilde{\Gamma}_{mp}^{\overline{bcd}} \partial_a 
+ \tilde{\Gamma}_{mp}^{\overline{acd}} \partial_b + \tilde{\Gamma}_{mp}^{\overline{bad}}  \partial_c \right. ~\nn
\\
&+& \left.
\tilde{\Gamma}_{mp}^{\overline{bca}} \partial_d \right] f_m ~.
\eea 
This is identical with Eq.~\eqref{curr_intrnsc_long} obtained from quantum kinetics, with a correction factor of $4$. The dissipationless part of Eq.~\eqref{energy_curr_third} is obtained to be
\bea \label{energy_dissipationless}
\bar \sigma_{a;bcd}^{{\rm H},\bar{\epsilon}} (\tau^0) &=& \sum_{p,m, {\bm k}} \dfrac{-1}{6\omega_{mp}^2} \left[ \tilde{\Gamma}_{mp}^{\overline{acd}} \partial_b + \tilde{\Gamma}_{mp}^{\overline{bad} } \partial_c 
+ \tilde{\Gamma}_{mp}^{\overline{bca}} \partial_d\right. \nn
\\
&-& \left. 3 \tilde{\Gamma}_{mp}^{\overline{bcd}} \partial_a \right] f_m ~.
\eea
This additional Hall conductivity does not have an analog in our quantum kinetic theory.

To summarize, by comparing the semiclassical wave-packet approach with the quantum kinetic theory, we conclude the following. For the cubic and quadratic scattering time-dependent contributions, both formalisms lead to similar results: i) the nonlinear Drude in cubic scattering [Eqs.~\eqref{NL_Drude} and \eqref{NL_Drude_SC}] and ii) Berry curvature quadrupole in quadratic scattering [Eqs.~\eqref{sigma_quad_tau} and \eqref{cond_quad_tau_BC}] time dependence. Only a trivial difference of factors arises due to the different choice of relaxation time in the semiclassical wave-packet approach and our quantum kinetic theory. For the linear-$\tau$ contributions, however, there are subtle differences. Both formalisms predict dissipative and dissipationless third-order currents. i) The Hall conductivities [Eqs.~\eqref{lin_tau_hall} and \eqref{cond_lin_tau_BC1}] are identical. However, the semiclassical wave-packet approach has an additional Hall contribution given in Eq. \eqref{cond_lin_tau_Hall_enrgy2} originating from the energy correction, which the quantum kinetics does not capture. ii) In the dissipative sector, both formalism gives similar results [Eqs.~\eqref{cond_lin_tau_Ohm} and \eqref{cond_lin_tau_Ohm_enrgy2}]. Finally, in the intrinsic third-order current, we find similarities and differences. i) The symplectic connection contribution to the dissipationless current given in Eq.~\eqref{sigma_sc_Hall} and Eq.~\eqref{curr_intrinsic_Xiang23} is captured by both formalisms. ii) The wave-packet approach does not capture the Berry curvature contribution to the third-order intrinsic dissipationless Hall current given in Eq.~\eqref{sigma_bc_sym}. iii) The dissipative conductivity originating from the energy correction of the wave packet given in Eq.~\eqref{energy_dissipative} is smaller by a factor of $1/4$ from the dissipative conductivity obtained from the quantum kinetic theory given in Eq.~\eqref{curr_intrnsc_long}. iv) The energy correction in the semiclassical theory [specified by Eq.~\eqref{energy_dissipationless}] also gives rise to a dissipationless contribution presented in Eq.~\eqref{energy_dissipationless}. This contribution is not present in the quantum kinetic theory.


\begin{table*}[t!]
    \centering
    \begin{ruledtabular}
    \caption{The classification of 122 magnetic point groups 
    based on different contributions of the dissipative (or Ohmic), and dissipationless (or Hall) third-order conductivity being finite in a crystalline system. 
    \label{table_symmetry_mpg_elements_Jahn}}
    \begin{tabular}{ccccc}
    \multirow{2}{*}{Magnetic point groups} & \multicolumn{2}{c}{$\rm Ohmic$} & \multicolumn{2}{c}{$\rm Hall$}
    \\
    \cline{2-5}
    & $\rm {\mathcal T}-even $ & $\rm {\mathcal T}-odd $ & $\rm {\mathcal T}-even $ & $\rm {\mathcal T}-odd $
    \\
    \hline
    {\bf R1}: (57) \\ 
    $1$, $-1$, $2$, $2^\prime$, $m$, $m^\prime$, $2/m$, $2^\prime/m^\prime$, $222$, $2^\prime2^\prime 2$, $mm2$, $m^\prime m2^\prime$, $m^\prime m^\prime 2$, & \multirow{5}{*}{\c} & \multirow{5}{*}{\c} & \multirow{5}{*}{\c} & \multirow{5}{*}{\c} 
    \\
    $mmm$, $m^\prime m^\prime m$, $4$, $4^\prime$, $-4$, $-4^\prime$, $4/m$, $4^\prime/m$, $422$, $4^\prime 22^\prime$, $42^\prime 2^\prime$, $4mm$, & & & & 
    \\
    $4^\prime m^\prime m$, $4m^\prime m^\prime$, $-42m$, $-4^\prime 2^\prime m$, $-4^\prime 2m^\prime$, $-42^\prime m^\prime$, $4/mmm$, $4^\prime/mm^\prime m$, & & & & 
    \\
    $4/mm^\prime m^\prime$, $3$, $-3$, $32$, $32^\prime$, $3m$, $3m^\prime$, $-3m$, $-3m^\prime$, $6$, $6^\prime$, $-6$, $-6^\prime$, $6/m$, & & & & 
    \\
    $6^\prime/m^\prime$, $622$, $6^\prime 22^\prime$, $6mm$, $6^\prime mm^\prime$, $-6m2$, $6/mmm$, $6^\prime/m^\prime mm^\prime$, $23$, $m-3$ & & & &  
    \\ [1ex]
    \hline
    {\bf R2}: (4) \\
    $62^\prime 2^\prime$, $6m^\prime m^\prime$, $-6m^\prime 2^\prime$, $6/mm^\prime m^\prime$ & \c & \x & \c & \c  
    \\ [1ex]
    \hline
    {\bf R3}: (50) \\
    $1^\prime$, $-11^\prime$, $-1^\prime$, $21^\prime$, $m1^\prime$, $2/m1^\prime$, $2^\prime/m$, $2/m^\prime$, $2221^\prime$, $mm21^\prime$, & \multirow{6}{*}{\c} & \multirow{6}{*}{\x} & \multirow{6}{*}{\c} & \multirow{6}{*}{\x} 
    \\
    $mmm1^\prime$, $m^\prime mm$, $m^\prime m^\prime m^\prime$, $41^\prime$, $-41^\prime$, $4/m1^\prime$, $4/m^\prime$, $4^\prime/m^\prime$, $4221^\prime$, & & & & 
    \\
    $4mm1^\prime$, $-42m1^\prime$, $4/mmm1^\prime$, $4/m^\prime mm$, $4^\prime/m^\prime m^\prime m$, $4/m^\prime m^\prime m^\prime$, & & & &
    \\
    $31^\prime$, $-31^\prime$, $-3^\prime$, $321^\prime$, $3m1^\prime$, $-3m1^\prime$, $-3^\prime m$, $-3^\prime m^\prime$, $61^\prime$, $-61^\prime$, & & & & 
    \\
    $6/m1^\prime$, $6^\prime/m$, $6/m^\prime$, $6221^\prime$, $6mm1^\prime$, $-6m21^\prime$, $-6^\prime m^\prime 2$, $-6^\prime m2^\prime$, & & & &
    \\
    $6/mmm1^\prime$, $6/m^\prime mm$, $6^\prime/mmm^\prime$, $6/m^\prime m^\prime m^\prime$, $231^\prime$, $m-31^\prime$, $m^\prime-3^\prime$ & & & &
    \\ [1ex]
    \hline
    {\bf R4}: (3) \\
    $4^\prime 32^\prime$, $-4^\prime 3m^\prime$, $m-3m^\prime$ & \c & \x & \x & \c
    \\ [1ex]
    \hline
    {\bf R5}: (3) \\
    $432$, $-43m$, $m-3m$ & \c & \c & \x & \x 
    \\
    \hline
    {\bf R6}: (5) \\
    $4321^\prime$, $-43m1^\prime$, $m-3m1^\prime$, $m^\prime-3^\prime m$, $m^\prime-3^\prime m^\prime$ & \c & \x & \x & \x  
    \\
\end{tabular}
\end{ruledtabular}
\end{table*}

\section{Symmetry analysis\label{symmetry}}
In this section, we present a crystalline symmetry analysis of both the dissipative and dissipationless third-order conductivities. Using Jahn's symbols for third-order conductivity in the {MTENSOR} program of the Bilbao crystallographic server~\cite{Gallego2019_actacrystal_symmetry_bilbao}, 
we determine the magnetic point groups that support these responses. Our analysis allows for the identification 
of materials capable of hosting one or more of these responses.

\begin{table}[h!]
    \centering
    \begin{ruledtabular}
    \caption{The classification of 90 magnetic point groups (type-I and type-III) based on the different components of the field-symmetric intrinsic 
    third-order conductivity being finite in a planar measurement setup. Type-II point groups do not allow intrinsic conductivities to be finite since they contain ${\mathcal T}$ as a symmetry element of the crystal.\label{table_symmetry_mpg_elements_planar}}
    \begin{tabular}{ccccccc}
    \multirow{2}{*}{Magnetic point groups} & $\bar{\sigma}_{x;xxx}$ & $\bar{\sigma}_{y;yyy}$ & \multicolumn{2}{c}{$\bar{\sigma}_{y;xxx}$} & \multicolumn{2}{c}{$\bar{\sigma}_{x;yyy}$}
    \\
    \cline{2-7}
    & $\rm O$ & $\rm O$ & $\rm O$ & $\rm H$ & $\rm O$ & $\rm H$
    \\
    \hline
    {\bf R1}: (33) \\ 
    $-1^\prime$, $2^\prime/m$, $2/m^\prime$, $m^\prime m2^\prime$, $m^\prime mm$, & \multirow{9}{*}{\x} & \multirow{9}{*}{\x} & \multirow{9}{*}{\x} & \multirow{9}{*}{\x} & \multirow{9}{*}{\x} & \multirow{9}{*}{\x}
    \\
    $m^\prime m^\prime m^\prime$, $4/m^\prime$, $4^\prime/m^\prime$, $4/m^\prime mm$, & & & & & &
    \\
    $4^\prime/m^\prime m^\prime m$, $4/m^\prime m^\prime m^\prime$, $-3^\prime$, & & & & & &
    \\
    $-3^\prime m$, $-3^\prime m^\prime$, $6^\prime$, $-6^\prime$, $6^\prime/m$, & & & & & & 
    \\
    $6/m^\prime$, $6^\prime/m^\prime$, $6^\prime 22^\prime$, $6^\prime m m^\prime$, & & & & & &
    \\
    $-6^\prime m^\prime 2$, $-6^\prime m2^\prime$, $6/m^\prime mm$, & & & & & &
    \\
    $6^\prime/mmm^\prime$, $6^\prime/m^\prime mm^\prime$, & & & & & &
    \\
    $6/m^\prime m^\prime m^\prime$, $m^\prime-3^\prime$, $4^\prime 32^\prime$, & & & & & &
    \\
    $-4^\prime 3m^\prime$, $m^\prime-3^\prime m$, $m-3m^\prime$, & & & & & & 
    \\
    $m^\prime-3^\prime m^\prime$ & & & & & &
    \\[1ex]
    \hline
    {\bf R2}: (7) \\
    $32^\prime$, $3m^\prime$, $-3m^\prime$, $62^\prime 2^\prime$, & \multirow{2}{*}{\x} & \multirow{2}{*}{\x} & \multirow{2}{*}{\x} & \multirow{2}{*}{\c} & \multirow{2}{*}{\x} & \multirow{2}{*}{\c}
    \\
    $6m^\prime m^\prime$, $-6m^\prime 2^\prime$, $6/mm^\prime m^\prime$ & & & & & &  
    \\ [1ex]
    \hline
    {\bf R3}: (1) \\
    $-42m$ & \x & \x & \c & \x & \c & \x
    \\ [1ex]
    \hline
    {\bf R4}: (22) \\
    $2$, $m$, $2/m$, $222$, $mm2$, $mmm$, & \multirow{5}{*}{\c} & \multirow{5}{*}{\c} & \multirow{5}{*}{\x} & \multirow{5}{*}{\x} & \multirow{5}{*}{\x} & \multirow{5}{*}{\x}
    \\
    $422$, $4^\prime 22^\prime$, $4mm$, $-4^\prime 2m^\prime$, & & & & & &
    \\
    $4/mmm$, $32$, $3m$, $-3m$, $622$, & & & & & & 
    \\ 
    $-6m2$, $6/mmm$, $23$, $m-3$, & & & & & & 
    \\
    $432$, $-43m$, $m-3m$ & & & & & &
    \\ [1ex]
    \hline
    {\bf R5}: (13) \\
    $2^\prime$, $m^\prime$, $2^\prime/m^\prime$, $2^\prime2^\prime 2$, $m^\prime m^\prime 2$, & \multirow{4}{*}{\x} & \multirow{4}{*}{\x} & \multirow{4}{*}{\c} & \multirow{4}{*}{\c} & \multirow{4}{*}{\c} & \multirow{4}{*}{\c}
    \\
    $m^\prime m^\prime m$, $42^\prime 2^\prime$, $4^\prime m^\prime m$, $4m^\prime m^\prime$, & & & & & &
    \\
    $-4^\prime 2^\prime m$, $-42^\prime m^\prime$, $4^\prime/mm^\prime m$, & & & & & &
    \\
    $4/mm^\prime m^\prime$ & & & & & &
    \\
    \hline
    {\bf R6}: (6) \\
    $3$, $-3$, $6$, $-6$, $6/m$, $6mm$ & \c & \c & \x & \c & \x & \c
    \\ [1ex]
    \hline
    {\bf R7}: (8) \\
    $1$, $-1$, $4$, $4^\prime$, $-4$, $-4^\prime$, $4/m$, & \multirow{2}{*}{\c} & \multirow{2}{*}{\c} & \multirow{2}{*}{\c} & \multirow{2}{*}{\c} & \multirow{2}{*}{\c} & \multirow{2}{*}{\c}
    \\
     $4^\prime/m$ & & & & & &
    \\
\end{tabular}
\end{ruledtabular}
\end{table}

In three dimensions, 122 magnetic point groups (MPGs) can describe magnetic systems with two possible spin-states (spin-orbit coupling not included). Among them, 32 represent classical point groups (group notation ${\mathcal R}$) describing nonmagnetic systems (atoms have zero magnetic moments) or magnetic states where neither time reversal (${\mathcal T}$) nor ${\mathcal R} {\mathcal T}$ is a symmetry operation. These are called colorless groups and are classified as type I. 
Another 32 point groups support classical point groups and time-reversal operations. These are called grey point groups, and they describe paramagnetic and diamagnetic systems preserving TR symmetry~[\onlinecite{Birss1963,Zhang2023_symmetry}]. They are classified as type-II groups. Since they are time-reversal symmetric, they do not support any intrinsic third-order response. The remaining 58 groups correspond to TR-broken magnetic systems like ferromagnets, antiferromagnets, and ferrimagnets. They are called the black and white point groups, classified as type III, and denoted by the prime of some crystalline symmetry element. Here, we will examine the 122 MPGs for the for the third-order conductivity. As evident from the above discussion, intrinsic and $\tau^2$ third-order conductivities are prohibited by the 32 type-II MPGs, while $\tau$ and $\tau^3$ extrinsic conductivities have no such symmetry constraint.

We will first illustrate Jahn's symbol for an arbitrary rank tensor. A rank-$N$ tensor using Jahn's symbol is written as ``$VN$". Symmetry or antisymmetry between specific spatial indices is indicated by parenthesis ``[]" or ``\{\}," respectively. For instance, a rank-4 tensor symmetric in the last three indices is represented as ``$V[V3]$". Two aspects have to be considered to determine the point group symmetry response of a tensor:  i) whether the tensor is polar/axial, and ii) whether it is ${\mathcal T}$-even/${\mathcal T}$-odd. For example, third-order conductivities exhibiting even powers in scattering time ($\tau^0$, $\tau^2$) are ${\mathcal T}$-odd, while those with odd powers in scattering time ($\tau$, $\tau^3$) are ${\mathcal T}$-even. The Jahn symbol for a ${\mathcal T}$-odd (${\mathcal T}$-even) rank-N tensor is denoted as ``$aVN$" (``$VN$"), respectively. Similarly, the axial (polar) nature of a rank-$N$ tensor is indicated by ``$eVN$" (``$VN$"). Following these notations, we construct the Jahn symbols for the third-order conductivity and predict their symmetry properties under the 122 magnetic point groups. 

The dissipative third-order conductivity $\bar{\sigma}^{\rm O}$ is fully symmetric in the spatial indices. Hence, the Jahn symbols for the ${\mathcal T}$-even (${ \mathcal T}$-odd) component of the dissipative conductivity are represented as $[V4]$ ($a[V4]$). For the dissipationless third-order conductivity, we follow a two-step procedure. First, we consider a general third-order conductivity tensor with field index symmetry. Using the Jahn symbols for such ${\mathcal T}$-even (${\mathcal T}$-odd) contributions to the conductivity as $V[V3]$ ($aV[V3]$), we obtain the different elements of the conductivity tensor. As   $\bar{\sigma}^{\rm H}$ can be obtained by subtracting the dissipative part from the field-symmetric conductivity, we calculate the elements of the Hall conductivity using $V[V3]-[V4]$. We separately obtain the finite elements of the $V[V3]$ and $[V4]$ tensors from the Bilbao server and then obtain the Hall conductivity tensor by subtracting them.

We summarize our symmetry analysis in  Table~\ref{table_symmetry_mpg_elements_Jahn}, where we have classified the 122 MPGs based on the finite or vanishing responses of the ${\mathcal T}$-even/${\mathcal T}$-odd dissipative and dissipationless conductivity under each point group symmetry transformations. We find that 57 MPGs allow both the dissipative and dissipationless responses, as shown in the row R1. In the 4 MPGs in row R2, only the ${\mathcal T}$-odd dissipative conductivity is forbidden, while in the 5 MPGs in row R6, only the ${\mathcal T}$-even dissipative conductivity is allowed by symmetry. Some of the MPGs (row R5), support only dissipative responses, while some (50 MPGs in row R3) allow only ${\mathcal T}$-even responses. Finally, in row R4, the 3 MPGs prohibit ${\mathcal T}$-even Hall and ${\mathcal T}$-odd Ohmic currents.

In addition, we have specifically considered the 
previously unexplored ${\mathcal T}$-odd longitudinal conductivity elements $\bar{\sigma}_{x;xxx}$, $\bar{\sigma}_{y;yyy}$, and the transverse components $\bar{\sigma}_{y;xxx}$ and $\bar{\sigma}_{x;yyy}$ in a planar set up, and presented a symmetry analysis in Table~\ref{table_symmetry_mpg_elements_planar}. Among these components, $\bar{\sigma}_{x;xxx}$ and $\bar{\sigma}_{y;yyy}$ allow for only dissipative (or Ohmic) contributions, while $\bar{\sigma}_{y;xxx}$ and $\bar{\sigma}_{x;yyy}$ allow for both the dissipative and non-dissipative (or Hall) contributions. We find that the longitudinal dissipative components $\bar{\sigma}_{x;xxx}$ and $\bar{\sigma}_{y;yyy}$ are allowed only for the 36 MPGs in row R4, R6, and R7. Similarly, the transverse responses $\bar{\sigma}_{y;xxx}$ and $\bar{\sigma}_{x;yyy}$ are allowed only in the 35 MPGs in row R2, R3, R5, R6, and R7. Among them, rows R2 and R6 cannot host the dissipative response from $\bar{\sigma}_{y;xxx}$ and $\bar{\sigma}_{x;yyy}$, while row R3 MPGs do not support their dissipationless counterparts. The 55 MPGs in rows R1 and R4 do not support any dissipative or dissipationless contribution from the components $\bar{\sigma}_{y;xxx}$ and $\bar{\sigma}_{x;yyy}$. Interestingly, the rows R5 and R7 allow all four components of the transverse response to be finite. Materials belonging to these MPGs can be potential candidates for exploring longitudinal and transverse third-order responses in the same setup.

\section{Intrinsic third order response in an antiferromagnetic monolayer}
\label{model_calc_tilted_warped}
In this section, we illustrate the existence of third-order intrinsic conductivities for a low-energy Hamiltonian that breaks TR symmetry. For that, we consider the low energy Hamiltonian up to $k^2$ around the $\Gamma$-point in the antiferromagnetic monolayer of SrMnBi$_2$. 
The model Hamiltonian can be expressed as [\onlinecite{Zhang2023_multipole}],
\be \label{eq:hp}
H_p ({\bm k} ) = \left( \frac{k^2}{2m^*} \right) \sigma_0 + v (\sigma_x k_y - \sigma_y k_x) + m (k_x^2-k_y^2)  \sigma_z ~.
\ee
Here, $\sigma_i(i=x,y,z)$ represents the Pauli matrices acting on the spin degrees of freedom, and $k=(k_x^2+k_y^2)^{1/2}$. 
The first two terms of the Hamiltonian represent a free electron gas with Rashba-type spin-orbit coupling. The last term represents the second-order band warping. The band warping term breaks the TR symmetry of the Hamiltonian, $\mathcal{T}H({\bm k})\mathcal{T}^{-1}\neq H(-{\bm k})$, where $\mathcal{T}=i\sigma_y K$ is the time reversal, and $K$ is the conjugation operator. This Hamiltonian has the $C_4 {\mathcal T}$ and $M_x {\mathcal T}$ symmetry and belongs to the symmetry group $4'm'm$, which allows third-order dissipationless responses to be finite as shown in R5 of Table~\ref{table_symmetry_mpg_elements_planar}. This has been earlier demonstrated in Ref.~\cite{Xiang2023}. Here, we focus particularly on the dissipative responses for this model Hamiltonian. 

Our analysis shows that the longitudinal dissipative responses ($\bar{\sigma}_{x;xxx}$, $\bar{\sigma}_{y;yyy}$) vanish for the low energy model described by Eq.~\eqref{eq:hp}, which is anticipated from the symmetries of the MPG (see row R5 of Table~\ref{table_symmetry_mpg_elements_planar}). Additionally, we find that the dissipative transverse responses ($\bar{\sigma}_{y;xxx}$, $\bar{\sigma}_{x;yyy}$) vanish for this low energy model, despite being allowed by the symmetries of the MPG. 
This highlights that `being allowed by point group symmetries' is a necessary condition for the responses to be finite and not a sufficient condition.
Breaking the system's crystalline symmetries can induce finite dissipative longitudinal and transverse responses. For example, adding strain in SrMnBi$_2$ makes both the dissipative intrinsic contributions finite, as shown below.

\begin{figure}
    \includegraphics[width=\linewidth]{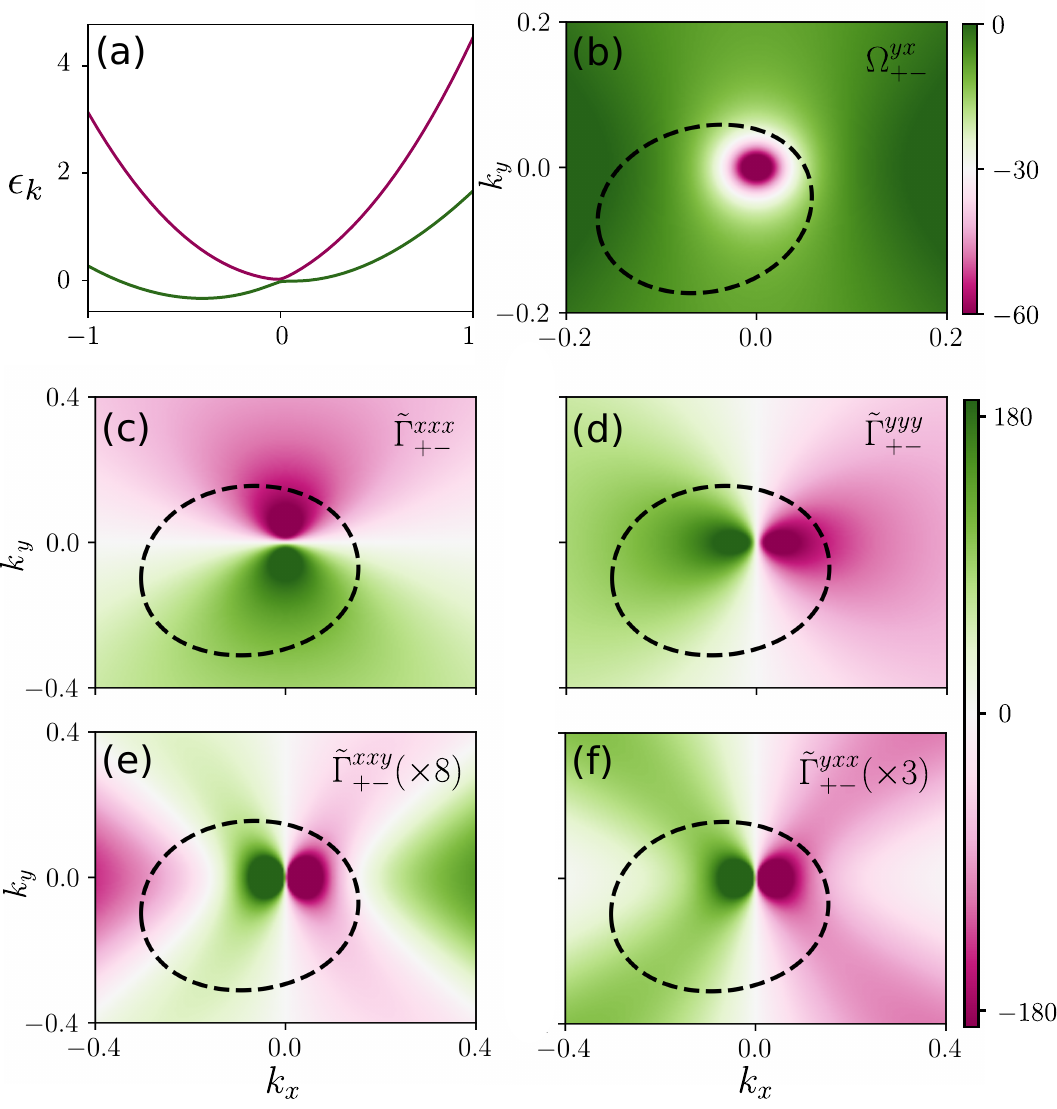}
    \caption{a) The band dispersion of the antiferromagnetic monolayer of SrMnBi$_2$ near the $\Gamma$-point, including the tilt and gap terms. 
    The magenta (green) curve represents the conduction (valence) band. b)-f) Momentum space distribution of the Berry curvature and different components of the symplectic connection. The dashed line (black) represents a Fermi surface contour. We used the following parameters:  $(m^\ast)^{-1}=2.4 ~\rm{eV}\cdot\AA^2$, $m=0.5 ~\rm{eV}\cdot\AA^2$, $v=0.1~ \rm{eV}\cdot\AA$, $\Delta=0~ \rm{eV}$, and $v_t=0.07~\rm{eV}\cdot\AA$.
    \label{fig2}}
\end{figure}

Including the symmetry-breaking terms, the total Hamiltonian of the strained system can be expressed as $H = H_p + H_t + H_\Delta$.  Here,  $H_p$ denotes the pristine system, $H_t$ captures the impact of strain via a tilt velocity term, and $H_\Delta$ denotes the gap opening term.
The symmetry breaking $H_t$, and the gap inducing $H_\Delta$ are specified by   
\be 
H_t = v_t (k_x + k_y) \sigma_0 ~,~~{\rm and}~~H_\Delta = \Delta \sigma_z~.
\ee
The addition of the tilt term breaks both the $M_x {\mathcal T}$ and $C_4 {\mathcal T}$ symmetry present in the Hamiltonian given in Eq.~\eqref{eq:hp} while the additional gap breaks the $C_4 {\mathcal T}$ symmetry keeping the $M_x {\mathcal T}$ symmetry intact. The energy dispersion for the total Hamiltonian is given by
\be \label{dsprsn}
\epsilon_k = \dfrac{k^2}{2m^\ast} + v_t (k_x + k_y) \pm \epsilon_{k0}~.
\ee
Here, $+(-)$ denotes the conduction (valence) band, and $\epsilon_{k0}$ is defined as $\epsilon_{k0}=\sqrt{v^2 k^2 + h_{+}^2 }$, where we have defined $h_{\pm}=[m (k_x^2 - k_y^2)\pm\Delta ]$. We present the energy dispersion in Fig.~\ref{fig2}(a). {The breaking of the symmetries can be easily verified from the dispersion. The breaking of $M_x {\mathcal T}$ symmetry can be easily verified as $\epsilon (k_x, k_y ) \neq \epsilon (-k_x, k_y )$. Furthermore, the breaking of the $C_4 {\mathcal T}$ symmetry can be checked from the dispersion as $\epsilon (-k_y, k_x ) \neq \epsilon (k_x, k_y )$.} To highlight the importance of the different terms in the Hamiltonian, we calculate the different band geometric quantities analytically. The BC for the Hamiltonian is given by
\be
\Omega_{+-}^{yx} =  \dfrac{ v^2h_{-}}{2 \epsilon_{k0}^{3} }~.
\ee
The different components of the SC are given by
\bea \label{geo_quant_model_warp}
\tilde{\Gamma}_{+-}^{yxx} &=& \dfrac{k_x v^2\left[m(h_{-})^2 -2mv^2k_y^2 -\Delta(v^2+4m\Delta)\right]}{2 \epsilon_{k0}^{5}} ~,~\nn
\\
\tilde{\Gamma}_{+-}^{xxy} &=&  \dfrac{k_x v^2h_{-}\left(v^2+2mh_{+}\right)}{4 \epsilon_{k0}^{5}} ~,~\nn
\\
\tilde{\Gamma}_{+-}^{xxx} &=& -  \dfrac{k_y m v^2}{2\epsilon_{k0}^{3}} ~~,~~~~{\rm and}~~~~ \tilde{\Gamma}_{+-}^{yyy} = -  \dfrac{k_x m v^2}{2\epsilon_{k0}^{3}} ~.
\eea
Note that all the band geometric quantities vanish if the band warping parameter $m$ and the gap $\Delta$ vanish in the system. We need either of them to be finite to have finite values of the band geometric quantities. Furthermore, as the tilt appears in the diagonal part of $H$, it does not affect the wavefunctions or the band geometric quantities. However, the presence of tilt velocity breaks the mirror symmetries, and it makes the Fermi surface asymmetric. We present the momentum space distribution of the BC in Fig.~\ref{fig2}(b) and different components of the SC in Fig.~\ref{fig2}(c)-(f). We find that the BC is peaked around the band edges, and the SC shows dipolar behavior. The dipolar behavior of SC can be understood by the fact that SCs can be expressed in terms of BC dipole, as $\partial_a \Omega^{bc}=2(\tilde{\Gamma}^{bac}-\tilde{\Gamma}^{cab})$. As the BC dipole typically has a dipolar structure, the SC also has a dipolar distribution.

\begin{figure*}[t!]
    \includegraphics[width=\linewidth]{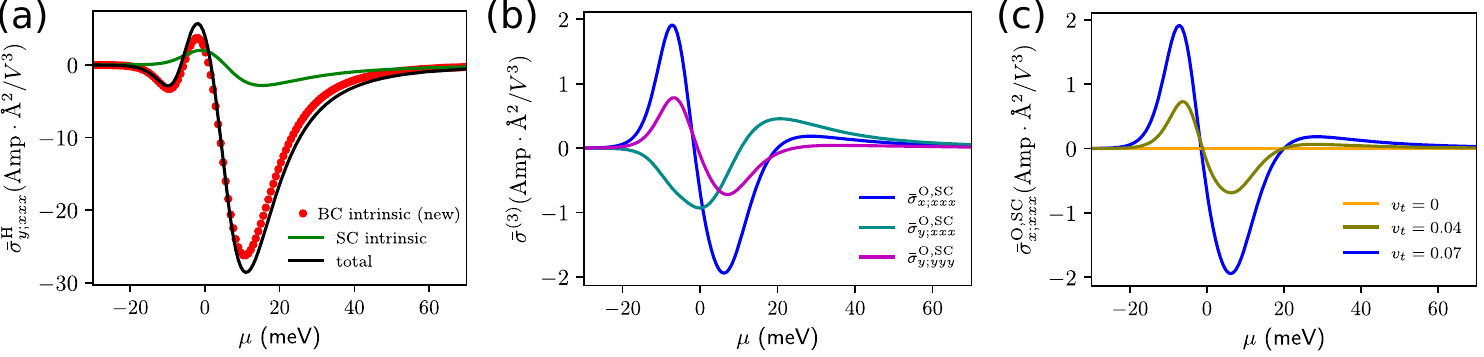}
    \caption{Intrinsic third order conductivities for an antiferromagnetic monolayer described by Eq.~\eqref{eq:hp}. a) The variation of the intrinsic (symmetrized) Hall conductivity with chemical potential ($\mu$). This dissipationless Hall contribution is finite in a pristine system. b) The dissipative longitudinal and transverse conductivity arising from the symplectic connection (SC) in the presence of a symmetry-breaking gap and tilt term. Both these responses vanish for the low energy model respecting all the point group symmetries. c) The variation of the longitudinal conductivity with $\mu$ for different values of the tilt velocity. We used the same parameters for our numerical calculations as those used in Fig.~\ref{fig2},  except for $\Delta=5$ meV. 
    \label{fig3}}
\end{figure*}

To illustrate the dissipationless and dissipative conductivities, we consider a planar setup where the electric field acts in the $x-y$ plane, ${\bm E}=(E_x, E_y, 0)$. Figure~\ref{fig3}(a) shows the third-order intrinsic Hall conductivity $\bar \sigma^{\rm H}_{y;xxx}$ as a function of the chemical potential. Both the BC and SC contribute to the intrinsic Hall current, although the impact of the BC contribution is dominant. Figure~\ref{fig3}(b) shows the third-order intrinsic dissipative conductivities as a function of the chemical potential. We find that the longitudinal dissipative responses, namely $\bar \sigma_{x;xxx}$ and $\bar \sigma_{y;yyy}$ vanish in the absence of tilt ($v_t \to 0$). To understand this, we note that the kernel of the longitudinal conductivity is proportional to $\tilde{\Gamma}_{xxx} v_x \partial_{\epsilon}f$ [see  Eq.~\eqref{curr_intrnsc_long}]. Now, for $v_t \to 0$, ${\tilde \Gamma}_{xxx}$ and the band velocity $v_x$ are odd functions of $k_y$ and $k_x$, respectively, and the electronic dispersion is an isotropic function. Consequently, we have $\tilde{\Gamma}_{xxx} v_x \partial_{\epsilon}f \propto k_x k_y$, and hence the longitudinal conductivity vanishes upon integration.
This highlights that the low energy model up to $k^2$ term given in Eq.~\eqref{eq:hp} does not capture the band anisotropy allowed by the point group symmetries. Adding a term like $H_t$ induces band anisotropy and breaks both the $M_x {\mathcal T}$ and the $C_4 {\mathcal T}$ symmetry, leading to a finite longitudinal response. This is verified in Fig.~\ref{fig3}(c), which presents the variation of the longitudinal dissipative contribution of $\bar \sigma_{x;xxx}$ for three different tilt velocities. 

As mentioned earlier, the dissipative transverse conductivity is symmetry-restricted. The integrand for $\bar \sigma^{\rm O}_{y;xxx}$, from Eq.~\eqref{curr_intrnsc_long} is obtained to be $\frac{2}{\omega_{mp}^2}(v_x \Gamma_{yxx} + 2 v_x \Gamma_{xxy} + v_y \Gamma_{xxx})\partial_\epsilon f$. Unlike longitudinal conductivity, each of these terms is an even function of momentum. Hence the transverse conductivity is not trivially zero. The integrand for $\bar \sigma^{\rm O}_{y;xxx}$, incorporating both the tilt and gap terms, using Eq.~\eqref{geo_quant_model_warp} can be expressed as 
\bea \label{kernal-a}  
\bar \sigma^{\rm O}_{y;xxx}({\bm k}) &=& \frac{m\hbar v^2}{{4\epsilon_{k0}^6}}\left[2m^2(k_x^4-k_y^4)+v^2(k_x^2-k_y^2) \right. \nn
\\
&+& \left. \epsilon_{k0}(k_x^2 -k_y^2)/m^* + 2\Delta m(k_x^2+k_y^2) \right. \nn
\\
 &+& \left.  v_t\epsilon_{k0}(k_x-k_y)\right]f'~.
\eea
Here, $f'$ is the energy derivative of the Fermi function. In the limit $\Delta \to 0, v_t \to 0$, the model Hamiltonian has $C_4 {\mathcal T}$-symmetry. Under this, the energy of the system given in Eq.~\eqref{dsprsn} remains unchanged. However, all the terms in the numerator of the right-hand side of Eq.~\eqref{kernal-a} change sign under $C_4 {\mathcal T}$ operation [$k_x \to - k_y$ and $k_y \to k_x]$. This forces the conductivity to vanish on integration over all ${\bm k}$-states consistent with our symmetry analysis. However, introducing a finite gap in the system breaks the $C_4 {\mathcal T}$ and makes $\bar \sigma^{\rm O}_{y;xxx}$ finite for $\Delta \neq 0$. Adding a tilt without a gap can not induce a transverse conductivity as the term involving tilt in Eq.~(57) is odd in $k_x$ and $k_y$. 


Note that the intrinsic conductivities are not particle-hole symmetric, as seen from the asymmetry of the response in conduction and valence band in Fig.~\ref{fig3}. This is expected, as the parabolic term, as well as the tilt term in the Hamiltonian, breaks the particle-hole symmetry. Being a Fermi surface property, all the intrinsic third-order conductivities vanish in the gap.

An important aspect of our work is that there is a large parameter regime, where the newly predicted contributions dominate the total third-order transport responses. To highlight this, we illustrate the dissipative and dissipationless conductivity in Fig.~\ref{fig_00}(a) and Fig.~\ref{fig_00}(b), categorizing them into `new' and previously `known' contributions. 
The shaded region in both plots marks the regime where the new contributions are the largest contributions to total third-order transport.



\section{scaling law for third order conductivities} \label{scaling_law}
Till now, we have focused on the third-order dc ($\omega=0$) conductivities. However, it is convenient in experiments to apply a frequency-dependent current and then measure the generated voltage, preferably via frequency-locked measurements. Motivated by this, we now discuss how the different conductivity contributions (specifically the intrinsic contributions) can be inferred from voltage measurements. Before proceeding, we will briefly summarize how the different harmonics are measured in transport experiments~\cite{Kang2019,QMa2019,Gao2023_science,wang_Nat2023_qun}.
In an experiment, the lock-in technique can distinguish nonlinear responses of various orders in the applied perturbation. A current with a frequency of a few tens to hundreds of Hertz is sent to the sample, and voltage with different frequencies is measured. For the transverse and longitudinal voltages, the voltage in the transverse ($V_\perp^{n \omega}$) and longitudinal ($V_\parallel^{n \omega}$) direction is measured. Additionally, we can use circular disc geometry to measure the angular dependence of different Hall effects and probe the impact of crystalline symmetry of the system~\cite{Kang2019, Lai2021,wang_Nat2023_qun}.  

In current-induced voltage measurements, both intrinsic and scattering time-dependent conductivities will contribute to the third-order voltage. Therefore, the scaling relations for the nonlinear voltage become vital for discerning the distinct conductivity contributions. As discussed earlier, the third-order conductivity has contributions that scale as $\tau^3$, $\tau^2$, and $\tau$ in addition to the intrinsic effects.  To distinguish between the different contributions, the measured longitudinal conductivity in the linear response regime can be used as a proxy for the scattering time ($\sigma_{xx} \propto \tau$). Experimentally, the scattering time can be varied by changing the temperature or other system parameters such as strain, sample thickness, or gate voltages. Such an approach has been used to separate different scattering time dependences in the nonreciprocal second-order transport~\cite{Kang2019,Sinha2022,Gao2023_science,wang_Nat2023_qun} as well as in the recently measured third-order responses~\cite{Lai2021,Wang2022}. Following Ref.~\cite{Kang2019}, with $V^{3\omega}_{\perp} = \rho j_{\perp}^{3 \omega}$, where $\rho$ is the linear resistivity, we can express,  
\be 
\frac{V^{3\omega}_\perp}{(V_\parallel)^3} = \frac{\sigma_{abcd}^{\perp}}{\sigma_{\parallel}}~.
\ee
In various order of scattering time we can express the transverse conductivity $\sigma^{(3)}_\perp$ as a function of $\sigma_{xx}$, $\sigma_\perp^{(3)}=\eta_{3\perp}\sigma_{xx}^3+\eta_{2 \perp}\sigma_{xx}^2+\eta_{1\perp}\sigma_{xx}+\eta_{0\perp}$. Here, the different powers of $\sigma_{xx}$ capture the different powers of the scattering time. For example, the first term represents the cubic scattering time. Note that although the cubic scattering time dependence does not have any dissipationless response, in real systems, band anisotropy can lead to some transverse response of a dissipative nature. Using this, the scaling law becomes 
\be \label{hall_scaling}
\frac{V^{3\omega}_\perp}{(V_\parallel)^3} = \eta_{3\perp} \sigma_{xx}^2 + \eta_{2\perp} \sigma_{xx} + \eta_{1\perp}  + \frac{\eta_{0\perp}}{\sigma_{xx}}~.
\ee
Note that a restricted version of this scaling law was used for third-order nonlinear conductivity in Ref.~\cite{Lai2021,Wang2022}, where only the second and third terms on the right-hand side were retained to account for the presence of time-reversal symmetry. We have generalized the third-order voltage scaling relation in Eq.~\eqref{hall_scaling} to include magnetic materials.

The scaling of the third-order dissipative response differs slightly from the transverse Hall response scaling law given in Eq.~\eqref{hall_scaling}. Since the quadratic scattering time dependence does not have a dissipative contribution, the scaling law for the longitudinal voltage becomes 
\be
\frac{V^{3\omega}_\parallel}{(V_\parallel)^3} = \eta_{3\parallel} \sigma_{xx}^2 + \eta_{1\parallel}  + \frac{\eta_{0\parallel}}{\sigma_{xx}}~.
\ee
Another simpler approach to finding the intrinsic part of the conductivity is to extract the third-order conductivity from the measured voltage. By experimentally varying $\sigma_{xx}$ by temperature and measuring the scaling of $\sigma^{(3)}$ with $\sigma_{xx}$, we can effectively segregate the intrinsic and other extrinsic contributions. Specifically, the intercept along the y-axis of the $\sigma^{3} - \sigma_{xx}$ plot corresponds to $\eta_0$, which is proportional to the intrinsic third-order conductivity contribution. In scenarios where the intrinsic part dominates, the measured third-order conductivity becomes independent of the linear conductivity~\cite{Gao2023_science}.
\section{Discussion\label{discuss}}
Here, we highlight a few essential points about the third-order conductivity derived in this paper. Specifically, we mention some of the limitations of our work and discuss the similarities and differences in the results obtained from the semiclassical wave-packet approach and quantum kinetic theory. 

In this paper, we have neglected multiband contributions involving more than two bands. We did this to simplify analytical calculations and avoid highly complicated expressions with limited insights. 
In principle, one can include those terms by deriving some appropriate sum rules or numerically. Exploring the impact of these multiband contributions, particularly in real materials, can be a good endeavor for the near future. 

In our calculations, we pursue the adiabatic switching-on approximation, which straightforwardly implies the factor of $\tau/2$ and $\tau/3$ for the relaxation of second and third-order density matrices, respectively, as considered in Eq.~\eqref{recursive_dm}. We find that these different relaxations are significant and have immense consequences on the final results. To demonstrate this explicitly, we alternatively considered $\tau$ as the relaxation of all order of density matrices. See Sec.~S8 of the SM~\cite{Note1} for details. Proceeding with this choice, we find significant changes in the conductivities. For the scattering time-independent part, the conductivity originating from $\rho^{\rm odo}$ changes as $\sigma^{\rm int,odo}\to 2\sigma^{\rm int,odo}$, the conductivity originating from $\rho^{\rm doo}$ changes as $\sigma^{\rm int,doo}\to 2\sigma^{\rm int,doo}$, and the conductivity originating from $\rho^{\rm ooo}$  ($\sigma^{\rm int,ooo}$) remains unchanged. These modifications lead to the emergence of a non-vanishing Fermi sea current contribution, which is given by
\bea \label{sea_1}
\bar{\sigma}_{a;bcd}^{\rm Sea} (\tau^0) &=& \frac{e^4}{\hbar^3} \sum_{m,p,{\bm k}} \dfrac{1}{3\omega_{mp}^2} \left[ 3\mathcal{G}_{mp}^{ab}\Omega_{mp}^{cd} + \operatorname{Im}[D_{mp}^{badc}] \right. \nn
\\
&-& \left. \operatorname{Im}[(\mathcal{D}_{mp}^b \mathcal{R}_{mp}^a) (\mathcal{D}_{pm}^d \mathcal{R}_{pm}^c) \right] f_m \nn
\\ [0.5ex]
&+& (c,d,b) + (d,b,c) ~.
\eea
Here, $(b,c,d)$ represents cyclic permutation. Similarly, for the linear-$\tau$-dependent extrinsic conductivities, we also get an unphysical Fermi sea contribution. Here, conductivities originating from the diagonal density matrices $\rho^{\rm ddo}$, $\rho^{\rm dod}$ and $\rho^{\rm doo}$, changes as $\sigma^{\rm ddo}(\tau)\to \frac{1}{6}\sigma^{\rm ddo}(\tau)$, $\sigma^{\rm dod}(\tau)\to \frac{2}{3}\sigma^{\rm ddo}(\tau)$, and $\sigma^{\rm doo}(\tau)\to \frac{1}{3}\sigma^{\rm doo}(\tau)$. The conductivities from off-diagonal density matrices $\rho^{\rm odd}$ and $\rho^{\rm odo}$ are modified as $\sigma^{\rm odd}(\tau)\to \frac{3}{2}\sigma^{\rm odd}(\tau)$ and $\sigma^{\rm odo}(\tau)\to \frac{1}{2}\sigma^{\rm odo}(\tau)$. Consequently, an additional Fermi sea contribution arises in the linear in $\tau$ extrinsic current, given by,
\bea \label{sea_2}
\bar{\sigma}_{a;bcd}^{\rm Sea} (\tau) &=& - \tau \dfrac{e^4}{\hbar^3} \sum_{m,p}^{\bm k} \dfrac{1}{\omega_{mp}^2} \left[\Delta_{mp}^{ab} \mathcal{G}_{mp}^{cd} + \Delta_{mp}^{ac} \mathcal{G}_{mp}^{db} \right. \nn
\\
&+& \left. \Delta_{mp}^{ad} \mathcal{G}_{mp}^{bc} \right] f_m ~.
\eea
These conductivities given in Eq.~\eqref{sea_1} and \eqref{sea_2} result in both dissipationless and dissipative currents, which can be finite even within the bandgap. Although a dissipationless current in the gap is allowed (intrinsic anomalous Hall effect), having a dissipative response in the bandgap (with no density of states) is unphysical. Thus, a choice of the same relaxation time in all orders of scattering time is likely to be incorrect. 

In our work, although we highlight the similarities and differences between the results obtained from quantum kinetic theory and the semiclassical wave-packet approach, a complete reconciliation is pending. The complete reconciliation of both these approaches needs more work. Our weakly held belief is that the semiclassical theory is missing some terms, and there is scope for improvement in the quantum kinetic theory by adding the effects of the finite width of the wave packet for calculating the electrical responses.

In this work, the intrinsic Hall current arises solely from the band geometry of the Bloch wavefunctions. This is the dominant mechanism at zero temperature and in the clean limit, whereas the side-jump and skew-scattering mechanisms~[\onlinecite{CXiao2019,Du2021}] play a crucial role with increased impurity density. These phenomena influence and modify the extrinsic ($\tau$ dependent) and the intrinsic conductivities. Here, we have not included such disorder-induced phenomena in the third-order current. Estimating the corrections to the predicted intrinsic third-order responses induced by the side jump and skew-scattering contributions is another interesting direction to pursue. 

Finally, we would like to mention a subtle difference between the semiclassical wave-packet approach used in literature and our paper. In previous literature based on the semiclassical wave-packet approach~\cite{HLiu2022_BCP_prb,Roy2022,Nag2022}, a linear-$\tau$-dependent third-order conductivity in addition to Eq.~\eqref{cond_lin_tau_BC1} has been obtained as
\be \label{additional_cont}
\bar \sigma_{a;bcd}= \dfrac{\tau}{6} \sum_{m ,{\bm k}}\left[ v_a v_b G_m^{cd} + v_a v_d G_m^{bc} + v_a v_c G_m^{bd} \right] f_m^{\prime\prime}~.~
\ee
This is proportional to the quantum metric and dispersive velocities.
It originates from the non-equilibrium distribution function of the form $\delta g = \sum_{\alpha=1}^{\infty}(\tau{\bm E}\cdot\nabla_{\bm k})^\alpha f(\tilde \epsilon)$ where electric field modified energy $\tilde \epsilon$ appears in the Fermi-Dirac distribution function. However, considering the Boltzmann equation of the form
\be \label{wrong_relax}
\dot{\bm k}\cdot\nabla_{\bm k}g=-\frac{g-f( \epsilon)}{\tau}~,
\ee
and following the usual iterative approach of Boltzmann equation, we find a solution like $\delta g = \sum_{\alpha=1}^{\infty}(\tau{\bm E}\cdot\nabla_{\bm k})^\alpha f(\epsilon)$. On the right-hand side of the above equation, the non-equilibrium distribution function relaxes to the equilibrium Fermi Dirac distribution function. With this solution, we do not obtain the contribution of Eq.~\eqref{additional_cont}. Interestingly, such a contribution given in Eq.~\eqref{additional_cont} is neither captured by our quantum kinetic theory, improving the consistency between the two theories.

\section{Conclusion} 
\label{conclude}

To summarize, we present a comprehensive picture of all third-order charge transport responses (in the regime $\omega \tau \ll 1$), using the quantum kinetic theory framework. We predicted novel dissipationless Hall and dissipative longitudinal third-order currents induced by band geometric quantities in time-reversal symmetry-broken systems. 
We demonstrated that the intrinsic Hall conductivities depend on band geometric quantities such as symplectic connection and Berry curvature. Performing a detailed symmetry analysis, we identified the magnetic point groups that can host the predicted dissipative and novel Hall responses. We illustrated the existence of these unique conductivities in an antiferromagnetic monolayer of SrMnBi$_2$. Additionally, we highlight that even if some responses are not allowed by the point group symmetries of the system, they can be made finite by breaking the crystalline symmetries via strain or other means.

Our comprehensive analysis of all contributions to the third-order conductivity highlights the importance of choosing the appropriate scattering timescale within the relaxation time approximation for nonlinear responses. 
We unified our quantum kinetic calculations with the semiclassical wave-packet formalism by doing a detailed comparison. We identified subtle differences and discrepancies between results obtained from the semiclassical framework in different works and our results. 
By doing a detailed comparison and highlighting the subtle differences explicitly, our work provides a foundation for the community to address them collectively.

Our study of third-order charge transport can be generalized beyond charge response. There are already several generalizations of second-order charge transport framework to thermoelectric and thermal transport~\cite{Zeng2020,Varshney2022,Zhou2022,Varshney2023}, valley transport~\cite{das_arxiv2023_nonlinear}, and spin Edelstein effects~\cite{CXiao2022,Xu2021,Lihm2022,Hayami2022}. Extending our work to investigate third-order responses in thermal, thermoelectric, and electro-thermal transport promises additional insights and new phenomena. Exploration of third-order responses in spin and valley degrees of freedom is another exciting frontier for fundamental physics motivated by our findings. Furthermore, our study can be generalized to include the effect of magnetic field. Building on the present study on non-reciprocal magneto-transport~\cite{Huang_PRL2022_intrinsic_nphe,Lahiri2022_resistivity} and our study in this paper, third-order reciprocal magnetotransport can be a future avenue to explore.    


\section{Acknowledgement}
D. M. is supported by the Department of Physics, IIT Kanpur. D. M. is thankful to Harsh Varshney and Shibalik Lahiri for many useful discussions. S. S. thanks the MHRD, India, for funding through the Prime Minister’s Research Fellowship (PMRF). K. D. acknowledges Prof. Binghai Yan for useful discussions. K. D. was supported by the Weizmann Institute of Science and the Koshland Foundation.  

\appendix

\bibliography{Intrnsc_3rd.bib}

\end{document}